\documentclass[11]{article}

\usepackage{epsfig}
\usepackage{amsfonts}
\usepackage{amsmath}
\usepackage{enumerate}

\usepackage{epsf}

\makeatletter\renewcommand{\fps@figure}{htbp}\makeatother

\usepackage{graphicx}

\begin{document}




\null\vskip-24pt




\vskip0.3truecm

\begin{center}

\vskip 3truecm

{\Large\bf

Multi-trace quasi-primary fields of $\mathcal{N}=4$ SYM$_4$\\ from AdS $n$-point functions
}\\ 

\vskip 1.5truecm


{\large\bf L. Hoffmann} {\large\bf, L. Mesref }{\large\bf, A. Meziane}{\large\bf, W.  R\" uhl}\footnote{email: {\tt hoffmann, mesref, meziane, ruehl@physik.uni-kl.de}}

\vskip 1truecm


{\it Department of Physics, Theoretical Physics\\

University of Kaiserslautern, Postfach 3049 \\

67653 Kaiserslautern, Germany}\\

\end{center}
\vskip 1truecm
\centerline{\bf Abstract}
We develop a recursive algorithm for the investigation of infinite sequences of quasi-primary fields obtained from chiral primary operators (CPOs) $O^I_k(x)$ and eventually their derivatives by applying operator product expansions and singling out particular $SO(6)$ representations. We show that normal products of $O_2$ operators can, to leading order, be expressed in terms of projection operators on representations of $SO(20)$ and discuss intertwining operators for $SO(6)$ representations. Furthermore we derive $\mathcal{O}(\frac{1}{N^2})$ corrections to AdS/CFT $4$-point functions by graphical combinatorics and finally extract anomalous dimensions by applying the method of conformal partial wave analysis. We find infinite sequences of quasi-primary fields with vanishing anomalous dimensions and interpret them as $\frac{1}{2}$-BPS or $\frac{1}{4}$-BPS fields.



\newpage

\section{Multi-trace operators in $SYM_4$}

AdS/CFT correspondence \cite{malda} is at present the most prominent way of extracting information on the strong coupling behavior of certain gauge field theories. We study the superconformal theory SYM$_4$ with gauge group $SU(N)_G$, $\mathcal{N}=4$ extended supersymmetry and $SU(4)_R$ ($SO(6)_R$) as $R$-symmetry. At strong 't Hooft coupling $\lambda = g_{YM}^2N$ it is dual to type $I \negthickspace IB$ supergravity on an AdS$_5$ $\otimes$ S$_5$ background geometry. Duality is meant in a twofold sense: The generating functional for connected Green functions of SYM$_4$ is identical with the off-shell value of the action of type $I \negthickspace IB$ supergravity which is a duality in the sense of Fourier transformations in the field space, and it connects the domain of small coupling constants in one theory with large coupling constants in the other one as in lattice field theories or in models with Olive-Montonen duality.     

In this superconformal Yang-Mills theory SYM$_4$ there exists an infinite number of  ''chiral primary operators'' (CPOs) \cite{lee}
\begin{equation} \label{1.1}
O^I_k(x)= Tr\{\Phi^{(i_1}\Phi^{i_2}\Phi^{i_3}...\Phi^{i_k)}-\text{traces}\}
\end{equation}
formed from the scalar fields $\Phi$ in the gauge supermultiplet by a trace over the $SU(N)$ gauge labels. The index ''i'' is an $SO(6)_R$ vector label corresponding to the $SU(4)_R$ dominant weight $[0,1,0]$. In (\ref{1.1}) we have formed a traceless symmetric $\bf {R}_6$-tensor of rank $k$ whose dominant $SU(4)_R$ weight is $[0,k,0]$. The dimension of a representation $[r_1,r_2,r_3]$ is 
\begin{equation} \label{1.2}
D=\frac{1}{12}(r_1+1)(r_2+1)(r_3+1)(r_1+r_2+2)(r_2+r_3+2)(r_1+r_2+r_3+3)
\end{equation}
The multiindex $I$ in (\ref{1.1}) can be mapped onto natural numbers
\begin{equation*}
\{I\} \leftrightarrow \{n \in {\bf{N}}: \, 1 \leq n \leq D_k\}
\end{equation*} 
It is known that these CPOs (\ref{1.1}) generate short, namely $\frac{1}{2}$-BPS, multiplets and that consequently their conformal field dimension is ''protected'' (i.e. canonical)
\begin{equation} \label{1.3}
\text{dim} \, O^I_k = k.
\end{equation}
Moreover 
\begin{equation} \label{1.4}
\text{dim} \,\Phi^i = 1.
\end{equation} 
as for a free scalar field.

This protectedness is synonymous with the existence and nontriviality of the limit 
\begin{equation} \label{1.5}
O^I_k(x)= \underset{\substack{x_i \rightarrow x \\ (\text{all} \, i)}}{\text{lim}}[Tr\{:\Phi^{(i_1}(x_1)\Phi^{i_2}(x_2)...\Phi^{i_k)}(x_k):-\text{traces}\}]
\end{equation}
where the double dots mean subtraction of all lower dimensional operators contained in the operator product expansion of the $\Phi$'s. These lower dimensional operators would be singular in the limit (\ref{1.5}).

In this work we develop a recursive algorithm for the investigation of infinite sequences of quasi-primary fields obtained from the CPOs (\ref{1.1}) and (eventually) their derivatives
\begin{equation} \label{1.6}
\partial_{\mu_1}\partial_{\mu_2}...\partial_{\mu_l} O_k^I(x)
\end{equation}
as limits of split-point polynomials of the fields (\ref{1.1}), (\ref{1.6}). If the number of such fields (\ref{1.1}) and (\ref{1.6}) is $k$, then we will call them $k$-trace\footnote{They are in general termed multi-trace operators.} operator fields.
Protectedness of the CPOs derived from the shortening of the supermultiplets repeats itself if we consider the limit
\begin{equation} \label{1.7}
M^J_{k(k_1k_2...k_r)}(x)= \underset{\substack{x_i \rightarrow x \\ (\text{all} \, i)}}{\text{lim}} \{:\sum_{\{I_i\}} \gamma^{J,I_1 I_2 ...I_r}_{k, k_1 k_2 ...k_r} \prod_{i=1}^r O_{k_i}^{I_i}(x_i):\}
\end{equation}
and if the $r$ traceless symmetric $\bf{R}_6$-tensors $O_{k_i}^{I_i}$ couple to a new traceless symmetric  $\bf{R}_6$-tensor of rank $k$ via the coupling coefficients $\gamma$, so that
\begin{equation} \label{1.8}
\text{dim} M^J_k = k +2n = \sum_{i=1}^r k_i
\end{equation}
where $n$ is the number of the $SO(6)$ contractions. The statement that $M^J_k$ generates a short, again $\frac{1}{2}$-BPS, multiplet from which (\ref{1.8}) follows will be referred to as ''Skiba's theorem'' \cite{skiba}.

In our approach the fields (\ref{1.7}), (\ref{1.8}) will also appear as simplest case. However, we are also interested in multi-trace operators with anomalous dimensions. First we project the formal product of the CPOs on a representation of $SU(4)_R$ ($SO(6)_R$) $[r,s,t]$
\begin{equation} \label{1.9}
 :\sum_{\{I_i\}} \gamma^{J,I_1 I_2 ...I_m}_{[r,s,t], k_1 k_2 ...k_m} \prod_{i=1}^m O_{k_i}^{I_i}(x_i):
\end{equation}
In this split-point form the multi-local operator has the dimension
\begin{equation*} 
k = \sum_{i=1}^m k_i 
\end{equation*}
but there may be other multi-local operators with the same representation $[r,s,t]$ and dimension $k$. Only certain linear combinations of such ''almost degenerate'' multi-trace operators possess an anomalous dimension. On the other hand, the knowledge of the anomalous dimensions is needed before one can perform the local limit of the multi-local operators (\ref{1.9}). For instance in the case $m=2$ the limit
\begin{equation} \label{1.10}
\underset{x_1 \rightarrow x_2}{\text{lim}} (x_{12}^2)^{-\frac{1}{2}\eta}:\sum_{\substack{k_1, k_2 \\ k_1+k_2=k}} A_{k_1,k_2} \gamma^{J,I_1 I_2}_{[r,s,t], k_1 k_2} \prod_{i=1}^2 O^{I_i}_{k_i}(x_i):
\end{equation}
defines the anomalous dimension $\eta$ of the local double-trace operator if the limit exists and is nontrivial.
Thus a general multi-trace operator can be constructed by taking linear combinations and decomposing the limit into steps, each of which needs a regularization in general (''chain of fusions''). 
On the other hand it is obvious, that the technical problems are reduced if some of the intermediate steps result in a field operator with protected dimension, so that neither anomalous dimensions nor linear combinations (eigenvalues and eigenvectors of a certain matrix) have to be determined. It is known from simpler conformal field theories, that the determination of all anomalous dimensions of a multiplet of almost degenerate fields leads to a cohomology problem \cite{lang1}. In this paper we will avoid these complications.
Namely, we apply an operator product expansion $r$ times
\begin{equation*}
O_{2k} \times O_2 \longrightarrow O_{2k+2} + ..., \qquad 1 \leq k \leq r-1
\end{equation*}
where we single out $SO(6)_R$ representations (e.g. denoted by Young diagram partitions)
\begin{equation*}
Y_{(2k)} \times Y_{(2)} \longrightarrow Y_{(2k+2)} + ... .
\end{equation*}
Each field $O_{2k+2}$ is scalar, so that Skiba's theorem applies to it and the operator product is essentially a simple ''normal product''. At the $r$-th step we apply 
\begin{equation*}
O_{2r} \times O_2 \longrightarrow O_{2r+2} + ...
\end {equation*}
and let $O_{2r+2}$ have tensor rank $l$ and $SO(6)_R$ representations according to 
\begin{equation} \label{1.11}
Y_{(2r)} \times Y_{(2)} \longrightarrow Y_{(2r+2)} + Y_{(2r+1,1)} + Y_{(2r,2)} + ...
\end {equation}
Taking traces of $SO(6)_R$ tensors is excluded at each step. This eliminates singular terms in the operator product expansion and thus simplifies the conformal partial wave expansion considerably.

In terms of $SU(4)_R$ weights, the three representations on the r.h.s of (\ref{1.11}) are
\begin{equation} \label{1.12}
[0, 2r+2, 0], \quad [1, 2r, 1], \quad [2, 2r-2, 2]
\end{equation}
respectively. According to the representation theory of the superconformal group \cite{dob, and} $\mathcal{N}=4$ SYM$_4$ may prossess chiral primary operators of the type
\begin{align}
&\tfrac{1}{2}-\text{BPS}: \quad [0, q, 0], \quad \Delta=q \geq 2, \label{1.13} \\
&\tfrac{1}{4}-\text{BPS}: \quad [p, q, p], \quad \Delta=q+2p,\quad (p\geq 2), \label{1.14} \\
& \tfrac{1}{8}-\text{BPS}: \quad [p, q, p+2k], \quad \Delta=3k+2p+q, \quad (k\geq 1), \label{1.15}
\end{align}
In the case $p=1$ (\ref{1.14}), there is no scalar field. We recognize that the representations (\ref{1.12}) fit into (\ref{1.13}), (\ref{1.14}).

As we have argued above (\ref{1.7}), there are many multi-trace quasi-primary operators for a given $SU(4)_R$ representation with approximate dimension $\Delta$. Only a few belong to short (BPS) multiplets. Examples have been analyzed in \cite{ryz} in the weak coupling regime of $\mathcal{N}=4$ SYM$_4$ at order $\mathcal{O}(g^2)$, namely either for all $\Delta \leq 7$ and exact in its $N$ dependence or for all $\Delta$ up to leading order in $N$ for $N \rightarrow \infty$ (including $\mathcal{O}(\frac{1}{N})$). A particular $\frac{1}{4}$-BPS field for $[p,q,p]$ arbitrary has been constructed from a linear combination of double-trace and single-trace operators. The number of traces involved is minimal obviously. Contrary to this construction, we consider multi-trace operators with a maximal number of traces in which case linear combinations are not needed.
  
In section 2 we consider two-point functions of the fields $O_2$ and normal products of several fields $O_2$ at leading order, including the vector besides the scalar normal products. It is shown that these two-point functions can be expressed in terms of projection operators on representations of $SO(20)$ realized in tensor spaces of $({\bf{R}}_{20})^n_\otimes$. In section 3 we discuss intertwining operators for $SO(6)$ representations realized alternatively as ${\bf{R}}_{6}$ and ${\bf{R}}_{20}$ tensors and derive from them the projection operators on irreducible $SO(6)$ representations in $({\bf{R}}_{6})^{2n}_\otimes$ and $({\bf{R}}_{20})^{n}_\otimes$. The AdS/CFT results on four-point functions at order $\mathcal{O}(\frac{1}{N^2})$ known from \cite{arut1} which are written in terms of such projectors are summarized. In the subsequent section 4 we formulate the last step of the fusion process $r \rightarrow r+1$ in terms of $SO(20)$ representations $Y_{(r+1)}$, $Y_{(r,1)}$ at $\mathcal{O}(1)$. The $\mathcal{O}(\frac{1}{N^2})$ corrections to these results are derived by graphical combinatorics in section 5. In the final section 6 we apply conformal partial wave analysis to extract anomalous dimensions. We emphasize, that our results prove Skiba's theorem once again recursively and yield an infinite number of short, $\frac{1}{4}$-BPS, multiplet fields.   

\section{Two-point functions of multi-trace operators at leading $\frac{1}{N^2}$ order}
\setcounter{equation}{0}
From the normalization of the single-trace operators
\begin{equation} \label{2.1}
\langle O^{I_1}(x_1)  O^{I_2}(x_2) \rangle = \frac{\delta^{I_1 I_2}}{(x_{12}^2)^2}, \quad (x_{12} = x_1-x_2)
\end{equation}
we obtain
\begin{equation} \label{2.2}
\langle :\prod_{a=1}^r O^{I_a}(x_1^{(a)})::\prod_{b=1}^r O^{J_b}(x_2^{(b)}): \rangle = \sum_{s\in S_r} \prod_{a=1}^r \frac{\delta^{I_a J_{s(a)}}}{(x_{1}^{(a)}-x_2^{(s(a))})^4}+ \mathcal{O}(\frac{1}{N^2})
\end{equation}
where $S_r$ is the symmetric group of $r$ elements and we denote by double dots the subtraction of the singular terms in each operator product. If we perform the limit 
\begin{equation} \label{2.3}
x_1^{(a)} \rightarrow x_1, \quad  x_2^{(b)} \rightarrow x_2, \quad \forall \,a,b \in \{1, 2,...r\}
\end{equation}
we obtain the two-point function for the totally symmetric $\bf{R}_{20}$-tensor field of rank $r$
\begin{equation} \label{2.4}
 :\prod_{a=1}^r O^{I_a}(x_1):.
\end{equation}
This field is reducible into $SO(6)$ representations of dimension $D_r$, since
\begin{equation} \label{2.5}
\frac{1}{r!} \sum_{s \in S_r} \prod_{a=1}^r \delta^{I_a J_{s(a)}} = \sum_{D_r} P_{D_r}^{I_1 I_2...I_r, J_1 J_2...J_r}
\end{equation}
where $P_D$ is a projection operator on a representation of dimension $D$,with the property $P^2_D = P_D$, and for $r=2$ it is known that $D_r \in \{\bf{105}, \bf{84}, \bf{20}, \bf{1}\}$. In this case ''$\bf{20}$'' involves one and ''$\bf{1}$'' involves two contractions. For general $r$ the noncontracted representations are
\begin{figure}[htb]
\begin{centering}
\includegraphics[scale=0.6]{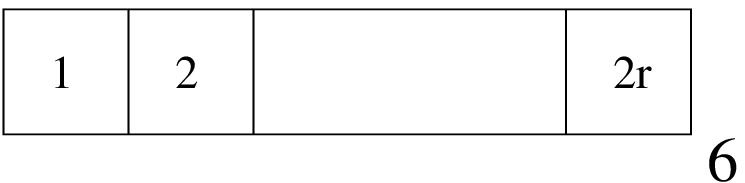}, \qquad \includegraphics[scale=0.6]{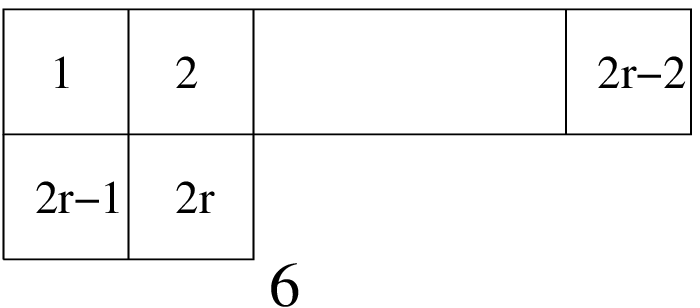}, \qquad \includegraphics[scale=0.6]{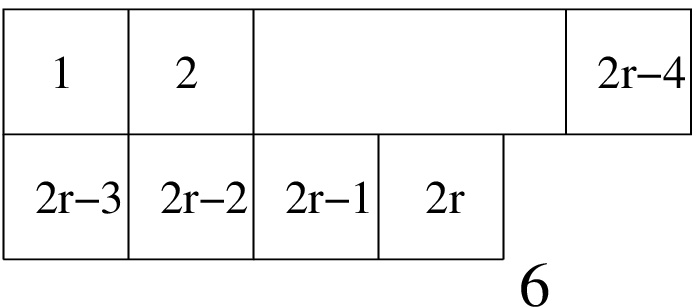}
\caption{Young tableau representations of $Y_{(2r)}$, $Y_{(2r-2,2)}$ and  $Y_{(2r-4,4)}$ }
\end{centering}
\end{figure}

\noindent
etc, but also three row Young tableaus appear.

In order to obtain $SO(6)$ tableaus with an odd number of blocks in the second row, we must start from ${\bf{R}}_{20}$-tensors with one block in the second row (see Figure 2).
\begin{figure}[htb]
\begin{centering}
\includegraphics[scale=0.6]{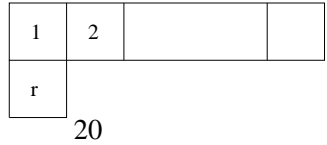} 
\caption{Representation of $Y_{r-1,1}$ }
\end{centering}
\end{figure}

\noindent
The symmetrizer $\Sigma_r$ and its transpose $\Sigma^T_r$ belonging to this tableau are
\begin{equation} \label{2.6}
e(r-1,1)= (e-(1,r)) \Sigma_{r-1}(1,...,r-1), \quad e^T(r-1,1) = \Sigma_{r-1}(1,...,r-1) (e-(1,r))
\end{equation}
where
\begin{equation} \label{2.7}
\Sigma_{r-1}(1,...,r-1) = \sum_{s \in S_{r-1}(1,...,r-1)} \negthickspace{\negthickspace{\negthickspace{\negthickspace{\negthickspace{\negthickspace s}}}}}
\end{equation}
is the total symmetrizer of the $r-1$ objects $(1,2,...,r-1)$. It is easy to see that (e.g. from (3.11))
\begin{equation} \label{2.8}
(e(r-1,1))^2 = n(r-1,1) e(r-1,1)
\end{equation}
with
\begin{equation} \label{2.9}
n(r-1,1) = r(r-2)!
\end{equation}

In order to produce a quasi-primary field with the symmetry $Y_{(r-1,1)}$, we apply the symmetrizer $e(r-1,1)$ (\ref{2.6}) to the right factor on the l.h.s. of (\ref{2.2}) and perform a Taylor expansion around a coordinate $x_2$.
\begin{multline} \label{2.10}
:e(r-1,1) \prod_{b=1}^r O^{J_b}(x_2^{(b)}):\,  = C_1 (x^{(r)}_2-x_2)^\mu :[O^{J_1}(x_2)\partial_\mu O^{J_r}(x_2)- O^{J_r}(x_2)\partial_\mu O^{J_1}(x_2)]\\ \prod_{b=2}^{r-1} O^{J_b}(x_2): + \mathcal{O}((x^r_2-x_2)^2)
\end{multline}
whereas for the left factor we need the transposed symmetrizer\footnote{See the discussion in the subsequent section.}
\begin{multline} \label{2.11}
:e^T(r-1,1) \prod_{a=1}^r O^{I_a}(x_1^{(a)}): \, = C_2 (x^{(r)}_1-x_1)^\mu :\sum_{a=1}^{r-1} [O^{I_a}(x_1)\partial_\mu O^{I_r}(x_1)- O^{I_r}(x_1)\partial_\mu O^{I_a}(x_1)]\\ \prod_{\substack{a'=1\\(a'\neq a)}}^{r-1} O^{I_{a'}}(x_1): + \mathcal{O}((x^r_1-x_1)^2)
\end{multline} 
Correspondingly the two-point function is
\begin{multline} \label{2.12}
(x^{(r)}_2-x_2)^\nu (x^{(r)}_1-x_1)^\mu  \sum_{A=1}^{r-1} \langle :[O^{I_A}(x_1)\partial_\mu O^{I_r}(x_1)-O^{I_r}(x_1)\partial_\mu O^{I_A}(x_1)] \prod_{\substack{a=1 \\ (a\neq A)}}^{r-1} O^{I_{a}}(x_1): \\ :[O^{J_1}(x_2)\partial_\nu O^{J_r}(x_2)-O^{J_r}(x_2)\partial_\nu O^{J_1}(x_2)] \prod_{b=2}^{r-1} O^{J_b}(x_2): \rangle
\end{multline}
and can be explicitly evaluated. 
With
\begin{align}
\Delta_1^{\mu\nu} &= \delta^{\mu\nu}-6\frac{x_{12}^\mu x_{12}^\nu}{x_{12}^2}, \label{2.13} \\
\Delta_2^{\mu\nu} &= \frac{x_{12}^\mu x_{12}^\nu}{x_{12}^2}, \label{2.14}
\end{align}
and
\begin{align}
&\partial_1^\mu \partial_2^\nu (x_{12}^2)^{-2} = 4 (x_{12}^2)^{-3} \Delta_1^{\mu\nu}, \label{2.15} \\
&\partial_{y_1}^\mu \partial_{y_2}^\nu ((y_1-x_2)^2 (x_1-y_2)^2)_{\arrowvert y_i = x_i}^{-2} = -16 (x_{12}^2)^{-5} \Delta_2^{\mu\nu} \label{2.16}
\end{align}
we obtain
\begin{align*} \label{2.17}
&(x_{12}^2)^{-2r-1} \sum_{A=1}^{r-1} 
\{4 \Delta_1^{\mu\nu} \delta^{I_r J_r} [\delta^{I_A J_1} \sum_{s_1 \in S_{r-2}} \prod_{\substack{a=1 \\ (a \neq A;\\ s(a) \neq 1)}}^{r-1} \delta^{I_a J_{s(a)}} \notag \\
&+\sum_{\substack{c,d=1 \\ (c \neq A, d \neq 1)}}^{r-1} \delta^{I_A J_d} \delta^{I_c J_1} \sum_{s_2 \in S_{r-3}} \prod_{\substack{a=1 \\ (a \neq A,c; \\ s(a) \neq 1,d)}}^{r-1} \delta^{I_a J_{s(a)}}]
-16 \Delta_2^{\mu\nu} [\delta^{I_A J_r} \delta^{I_r J_1} \sum_{s_1 \in S_{r-2}} \prod_{\substack{a=1 \\ (a \neq A;\\ s(a) \neq 1)}}^{r-1} \delta^{I_a J_{s(a)}} \notag \\
\end{align*}
\begin{align}
&+\sum_{\substack{c,d=1 \\(c \neq A,r, d \neq 1,r)}}^{r-1} \delta^{I_A J_1} \delta^{I_r J_d} \delta^{I_c J_r} +  \delta^{I_A J_d} \delta^{I_r J_1} \delta^{I_c J_r} +
 \delta^{I_A J_r} \delta^{I_r J_d} \delta^{I_c J_1} \sum_{s_2 \in S_{r-3}} \prod_{\substack{a=1 \\ (a \neq A,c,\\ s(a) \neq 1,d)}}^{r-1} \delta^{I_a J_{s(a)}} \notag \\
&+\sum_{c,d,e,f} \delta^{I_A J_d} \delta^{I_c J_1} \delta^{I_r J_e} \delta^{I_f J_r} \sum_{s_3 \in S_{r-4}} \prod_{\substack{a=1 \\ (a \neq A,c,f,\\ s(a) \neq 1,d,e)}}^{r-1} \delta^{I_a J_{s(a)}}] \notag \\
&-(I_A \leftrightarrow I_r)-(J_1 \leftrightarrow J_r) + (I_A \leftrightarrow I_r, J_1 \leftrightarrow J_r) \}
\end{align}
A general result of CFT says that a quasi-primary vector field of dimension $\delta = 2r+1$ ought to have a two-point function
\begin{equation} \label{2.18}
C (\delta^{\mu\nu}-2\frac{x_{12}^\mu x_{12}^\nu}{x_{12}^2}) (x_{12}^2)^{-\delta} P_Y^{I_1...J_r}
\end{equation}
In the cases $r=2$ and $r=3$ we have
\begin{equation} \label{2.19}
P_{Y_{(1,1)}}^{I_1 I_2, J_1 J_2} = \frac{1}{2}(\delta^{I_1 J_1} \delta^{I_2 J_2} - \delta^{I_1 J_2} \delta^{I_2 J_1})
\end{equation}
and
\begin{equation} \label{2.20}
P_{Y_{(2,1)}}^{I_1 I_2 I_3, J_1 J_2 J_3} = \frac{1}{3} [\delta^{I_3 J_3}(\delta^{I_1 J_2} \delta^{I_2 J_1} + \delta^{I_1 J_1} \delta^{I_2 J_2}) - \delta^{I_3 J_1}(\delta^{I_1 J_2} \delta^{I_2 J_3} + \delta^{I_2 J_2} \delta^{I_1 J_3})].
\end{equation}
The latter expression is symmetric under the transformations $I_1 \leftrightarrow I_2$ and antisymmetric under $J_1 \leftrightarrow J_3$ respectively.
For the cases $r=2$ and $r=3$ we obtain from (\ref{2.17}), that (\ref{2.18}) is valid with
\begin{equation} \label{2.21}
C = 8 r(r-2)!
\end{equation}
We conclude that the vector fields (\ref{2.10}) and its adjoint (\ref{2.11}) are indeed quasi-primary and belong to the irreducible representation $Y_{(r-1,1)}$ of $SO(20)$. Under $SO(6)$ it reduces to
\begin{equation} 
\sum_n Y_{(2r-2n-1,2n+1)} + ... \notag
\end{equation}
\begin{figure}[htb]
\begin{centering}
\includegraphics[scale=0.6]{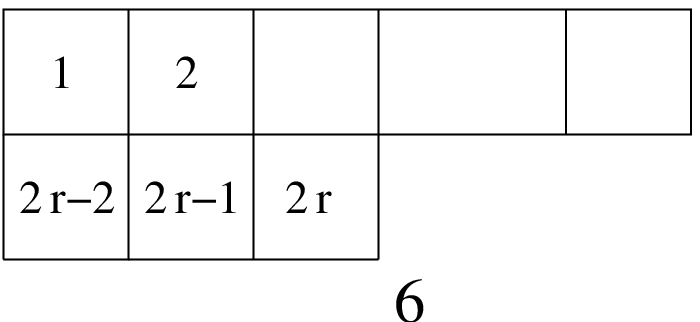} 
\caption{Young tableau of $Y_{(2r-2n-1,2n+1)}$ for $n=1$, representing  $Y_{(2r-3,3)}$  }
\end{centering}
\end{figure}
Moreover, the two point functions (\ref{2.2}), (\ref{2.17}) receive $\frac{1}{N^2}$ corrections from an AdS/CFT calculation which is different for the various $SO(6)$ representations. They will be derived in the sequel.

Thus we have found by construction of two-point functions that the fields (\ref{2.10}), (\ref{2.11}) are quasi-primary at order $\mathcal{O}(1)$, which is a nontrivial statement for composite fields containing derivatives. In section $6$ we will present the results, showing that for $SO(6)_R$ representations $Y_{(2r+1,1)}$ ($[1,2r,1]$ in $SU(4)_R$ Dynkin labels) and approximate dimensions $2r+2+l$, where $l$ is the space-time tensor rank and $l \geq 1$, the anomalous dimensions vanish at order $\mathcal{O}(\frac{1}{N^2})$. In the case $l=1$ these are the fields in (\ref{2.10}), (\ref{2.11}). We consider this as a strong hint that the fields are quasi-primary and not a mixture of different fields whose anomalous dimensions cancel each other accidentally. If this is true, then they belong to conformal multiplets and have minimal $Y$-charge in these supermultiplets. But since they are composite fields, we do not call them primary.

\noindent
For the composite fields with $SO(6)_R$ representations $Y_{(2r,2)}$ and tensor rank $l=0$ not treated in this section analogous formulae can be derived and the same statements concerning anomalous dimensions and quasi-primarity will follow from calculations performed in section $5$.

\section{Intertwining operators for $SO(6)$ representations in $\bf{R}_6$ and $\bf{R}_{20}$ tensor spaces}
\setcounter{equation}{0}
The mapping between symmetric traceless tensors from $({\bf{R}}_6)^2_\otimes$ into a subspace of ${\bf{R}}_{20}$ is achieved by intertwining matrices $C^I_{ij}$ defined by\footnote{We will never need the matrices $\{C^I_{ij}\}$ in explicit form.} \cite{lee}
\begin{equation} \label{3.1}
C^I_{ij}, \quad i, j \in \{1,2, ...,6\}, \quad I \in \{1,2, ...,20\}
\end{equation}
with the properties
\begin{align}
&C^I_{ij} = C^I_{ji}, \quad C^I_{ii} = 0, \label{3.2} \\
&Tr \, C^I C^J = \delta^{IJ} \label{3.3}
\end{align}
They satisfy the completeness relation
\begin{equation} \label{3.4}
\sum_I C^I_{i_1 i_2} C^I_{j_1 j_2}= \frac{1}{2}\{\delta^{i_1 j_1} \delta^{i_2 j_2} + \delta^{i_1 j_2} \delta^{i_2 j_1} - \frac{1}{3} \delta^{i_1 i_2} \delta^{j_1 j_2}\}
\end{equation}
The r.h.s. of (\ref{3.4}) defines the projector 
\begin{equation*} 
P_{20}^{i_1 i_2, j_1 j_2}
\end{equation*}
on the $20$-dimensional subspace of  $({\bf{R}_6})^2_\otimes$ carrying the representation $Y_{(2)}$ of $SO(6)$ (see Figure 4).
\begin{figure}[htb]
\begin{centering}
\includegraphics[scale=0.6]{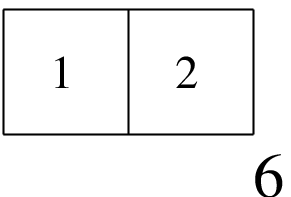} 
\caption{Young tableau for the representation $Y_{(2)}$ }
\end{centering}
\end{figure}

\noindent
Since only the subgroup $SO(6)$ of $SO(20)$ is relevant, there exist many other $SO(6)$ invariant tensors in $({\bf{R}_{20}})^m_\otimes$ besides Kronecker and Weyl tensors, namely all traces
\begin{equation} \label{3.5}
C^{I_1 I_2 ...I_m} = Tr \, (C^{I_1} C^{I_2}...C^{I_m}), \quad (m \geq 3)
\end{equation}
which are symmetric under cyclic permutations and reversing the order of the labels $\{I_1,...,I_m\}$. 
 Thus for any $m \geq 3$ there are 
\begin{equation*} 
\tfrac{1}{2}(m-1)!
\end{equation*}
different such invariant tensors.

Irreducible tensor representations of $SO(6)$ are conventionally realized on tensor subspaces  of $({\bf{R}_{6}})^n_\otimes$ characterized by a Young tableau $Y$ of $n$ blocks, the corresponding projector on the subspace is denoted by $P_Y$\footnote{One also uses $P_{D(Y)}$, where $D(Y)$ is the dimension of the representation.}. For a given tableau $Y$ one uses the symmetrizer
\begin{equation} \label{3.6}
e_Y = \mathcal{Q}_Y \mathcal{P}_Y
\end{equation}
where $\mathcal{P}_Y$ symmetrizes the rows and $\mathcal{Q}_Y$ antisymmetrizes the columns. Besides $e_Y$ we also need the transposed symmetrizer $e_Y^T$\footnote{E.g. for the definition of the adjoint field, see (\ref{2.11}).}. 
If we expand (\ref{3.6}), we obtain
\begin{equation} \label{3.7}
e_Y = \sum_{g \in S_n} \eta_g g
\end{equation}
with
\begin{equation} \label{3.8}
\eta_g \in \{-1,0,1\}
\end{equation}
so that $e_Y$ appears as an element of the group algebra of the symmetric group.
In turn we define
\begin{equation} \label{3.9}
e^T_Y = \sum_{g \in S_n} \eta_g g^{-1}
\end{equation}
As will be seen ((\ref{2.8}), (\ref{2.9}), (\ref{2.21})) the relation
\begin{equation} \label{3.10}
e^2_Y = n(Y) e_Y
\end{equation}
is used for normalizing the projector $P_Y$.
In fact $n(Y)$ can easily be calculated as
\begin{equation} \label{3.11}
n(Y)= \sum_{g \in S_n} \eta_g \eta_{g^{-1}}
\end{equation}
Then we can define the projector in components as
\begin{align} \label{3.12}
P_Y^{i_1 i_2 ...i_n, j_1 j_2 ...j_n} &= \frac{1}{n(Y)}\{e^T_Y(i) \prod_{k=1}^n \delta^{i_k j_k} - \text{traces}\} \\ \notag
				     &= \frac{1}{n(Y)}\{e_Y(j) \prod_{k=1}^n \delta^{i_k j_k} - \text{traces}\}	
\end{align} 
with the correct normalization
\begin{equation} \label{3.13}
P_Y^2 = P_Y
\end{equation}
The first term in the curly bracket of (\ref{3.12}) is called the ''no-trace term'', $P_{Y, \text{n-tr}}$. Then it is easy to see that (\ref{3.13}) is equivalent to
\begin{equation} \label{3.14}
(P_{Y, \text{n-tr}})^2 = P_{Y, \text{n-tr}}
\end{equation}
which can be shown to follow from (\ref{3.10}), (\ref{3.12}). The traces in the ''trace terms'' of  (\ref{3.12}) are taken in $\{i_1 i_2 ... i_n\}$ or  $\{j_1 j_2 ... j_n\}$ separately. 

Now we can define intertwining operators for each representation $Y$ of $SO(6)$ realized in $({\bf{R}_6})_{\otimes}^{2n}$ as a map into $({\bf{R}_{20}})_{\otimes}^{n}$ by the action of $e_Y$ on a product of $C$'s
\begin{align} 
Q_{Y,i_1 i_2 ... i_{2n}}^{I_1 I_2 ... I_n} &=  \frac{1}{n(Y)} e_Y(i) \{\prod_{l=1}^n C_{i_{2l-1} i_{2l}}^{I_l} - \text{traces}\}  \label{3.15} \\
					&= (\prod_{l=1}^n C_{j_{2l-1} j_{2l}}^{I_l}) P_Y^{j_1 j_2 ...j_{2l}, i_1 i_2 ...i_{2l}} \label{3.16}
\end{align}
where $Q_Y$ depends on $Y$ as well as on the ''pairing'' of the labels $\{i_1 i_2 ... i_{2l}\}$.
From (\ref{3.16}) it is obvious that we can also define a transposed intertwiner
\begin{equation} \label{3.17}
Q_{Y,i_1 i_2 ... i_{2n}}^{T, I_1 I_2 ... I_n} =  P_Y^{i_1 i_2 ...i_{2l}, j_1 j_2 ...j_{2l}} (\prod_{l=1}^n C_{j_{2l-1} j_{2l}}^{I_l})
\end{equation}
Contracting $Q_Y$ and $Q_Y^T$ on the $\bf{R}_6$ tensor labels, gives the projector on the representation space of $({\bf{R}_{20}})_{\otimes}^n$
\begin{equation} \label{3.18}
 P_Y^{(20) I_1 I_2 ...I_n, J_1 J_2 ...J_n} = \sum_{i} Q_{Y,i_1 i_2 ... i_{2n}}^{T, I_1 I_2 ... I_n} Q_{Y,i_1 i_2 ... i_{2n}}^{J_1 J_2 ... J_n}
\end{equation}
In turn, contracting them on the $\bf{R}_{20}$ tensor labels gives the paired projector $P_Y^{(6)}$
\begin{equation} \label{3.19}
P_Y^{(6)} \underset{\text{pairing}}{\rightarrow} P_Y^{(6) (i_1 i_2)...(i_{2n-1} i_{2n}), (j_1 j_2)...(j_{2n-1} j_{2n})} = \sum_{I} Q_{Y,i_1 i_2 ... i_{2n}}^{T, I_1 I_2 ... I_n} Q_{Y,j_1 j_2 ... j_{2n}}^{I_1 I_2 ... I_n}
\end{equation}
where the pairing corresponds to the operation
\begin{equation} \label{3.20}
 P_Y^{(6) ... (i_1 i_2)...} = \frac{1}{2}(\delta_{i_1 i'_1}\delta_{i_2 i'_2}+\delta_{i_1 i'_2}\delta_{i_2 i'_1}) P_Y^{(6) ... i'_1 i'_2...}
\end{equation}

Next we summarize some of the results that were originally derived by G. Arutyunov et al.\cite{arut2} for the calculation of the four-point function
\begin{equation} \label{3.21}
\langle O_2^{I_1}(x_1)O_2^{I_2}(x_2)O_2^{J_1}(x_3)O_2^{J_1}(x_4) \rangle
\end{equation}
from AdS/CFT at order $\frac{1}{N^2}$. The projectors in $({\bf{R}_{20}})_{\otimes}^{2}$ belonging to $SO(6)$ tensor representations of dimension $D$ that arise in (\ref{3.21}) are 
\begin{figure}[htb]
\begin{centering}
\includegraphics[scale=0.6]{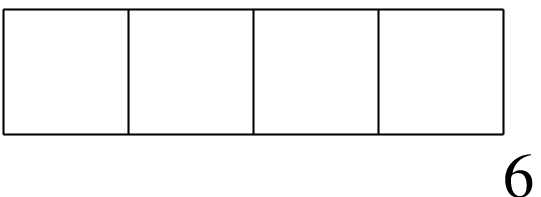}, \qquad \qquad \includegraphics[scale=0.6]{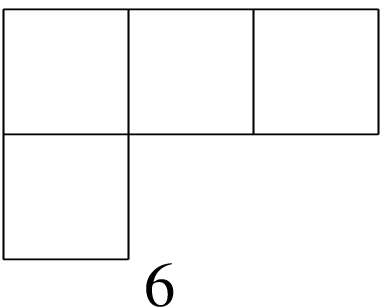}, \qquad \qquad  \includegraphics[scale=0.6]{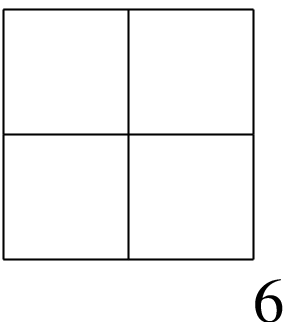} 
\caption{Representations $Y_{105}$,  $Y_{175}$,  $Y_{84}$ respectively }
\end{centering}
\end{figure}

\noindent
with their corresponding projectors
\begin{align} 
&P_{\bf{105}}^{I_1 I_2, J_1 J_2} = \frac{1}{6}(\delta_{I_1 J_1}\delta_{I_2 J_2}+\delta_{I_1 J_2}\delta_{I_2 J_1})+\frac{1}{60}\delta_{I_1 I_2}\delta_{J_1 J_2}+ \frac{2}{3}C^{I_1 J_1 I_2 J_2} -  \frac{4}{15}C_{+}^{I_1 I_2 J_1 J_2},  \label{3.22} \\
&P_{\bf{175}}^{I_1 I_2, J_1 J_2} = \frac{1}{2}(\delta_{I_1 J_1}\delta_{I_2 J_2} - \delta_{I_1 J_2}\delta_{I_2 J_1}) + \frac{1}{2} C_{-}^{I_1 I_2 J_1 J_2},  \label{3.23} \\
&P_{\bf{84}}^{I_1 I_2, J_1 J_2} = \frac{1}{3}(\delta_{I_1 J_1}\delta_{I_2 J_2}+ \delta_{I_1 J_2}\delta_{I_2 J_1}) + \frac{1}{30}\delta_{I_1 I_2}\delta_{J_1 J_2} - \frac{2}{3}C^{I_1 J_1 I_2 J_2} -  \frac{1}{3}C_{+}^{I_1 I_2 J_1 J_2}.  \label{3.24}
\end{align}

\noindent
as well as 
\begin{figure}[htb]
\begin{centering}
\includegraphics[scale=0.6]{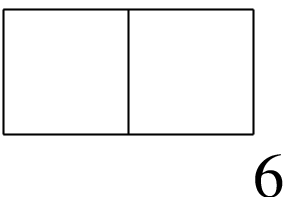}, \qquad \qquad \includegraphics[scale=0.6]{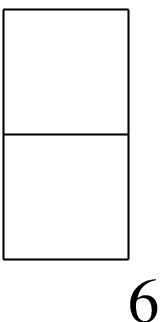} 
\caption{Representations $Y_{20}$ and $Y_{15}$ }
\end{centering}
\end{figure}

\noindent
with the projectors
\begin{align} 
&P_{{\bf{20}}}^{I_1 I_2, J_1 J_2} = \frac{3}{5}C_{+}^{I_1 I_2 J_1 J_2}-\frac{1}{10}\delta_{I_1 I_2}\delta_{J_1 J_2}, \label{3.25} \\
&P_{{\bf{15}}}^{I_1 I_2, J_1 J_2} = -2 C_{-}^{I_1 I_2 J_1 J_2} \label{3.26}
\end{align}
and finally the representation $Y_{1}$ with
\begin{equation} \label{3.27}
 P_{{\bf{1}}}^{I_1 I_2, J_1 J_2} = \frac{1}{20} \delta_{I_1 I_2}\delta_{J_1 J_2}
\end{equation}

Here we used the notations (\ref{3.5}) and
\begin{equation} \label{3.28}
C_{\pm}^{I_1 I_2 J_1 J_2} = \frac{1}{2} (C^{I_1 I_2 J_1 J_2} \pm C^{I_2 I_1 J_1 J_2})
\end{equation}
The projectors on the two $SO(20)$ representations in $({\bf{R}}_{20})^2_\otimes$ with the tableaus of Figure 7
\begin{figure}[htb]
\begin{centering}
\includegraphics[scale=0.6]{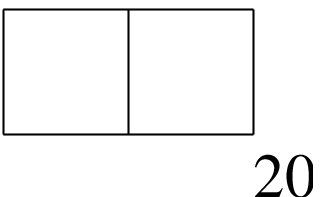} \quad \text{and} \quad \includegraphics[scale=0.6]{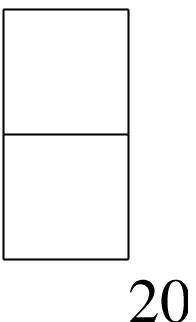} 
\caption{The tableaus have dimensions ${\bf{209}}$ and ${\bf{190}}$, respectively}
\end{centering}
\end{figure}
decompose into irreducible $SO(6)$ representations as follows
\begin{align} 
P^{(20)}_{{\bf{209}}} &= P^{(6)}_{{\bf{20}}}+P^{(6)}_{{\bf{84}}}+P^{(6)}_{{\bf{105}}} \label{3.29} \\
P^{(20)}_{{\bf{190}}} &= P^{(6)}_{{\bf{15}}}+P^{(6)}_{{\bf{175}}} \label{3.30}
\end{align}
For obvious reasons we call the representations of $SO(6)$ with $D \in \{\bf{1}, \bf{20}, \bf{84}, \bf{105}\}$ ''even'' and those with $D \in \{\bf{15}, \bf{175}\}$ ''odd''.

The AdS/CFT computation of (\ref{3.21}) up to order $\mathcal{O}(\frac{1}{N^2})$ results in \cite{arut1, eden, arut2}
\begin{multline} \label{3.31}
(x_{12}^2 x_{34}^2)^{-2} \{20  P_{1}^{I_1 I_2, J_1 J_2} + u^2 \sum_{D}  P_{D}^{I_1 I_2, J_1 J_2} + v^2 (\sum_{\text{even}\, D}-\sum_{\text{odd}\,  D}) P_{D}^{I_1 I_2, J_1 J_2} \\
+ \frac{1}{N^2}[\sum_{D} ( \lambda_D(u,v) \phi(u,v) + \mu_D(u,v))  P_{D}^{I_1 I_2, J_1 J_2}] + \mathcal{O}(\frac{1}{N^4})\}
\end{multline}
where
\begin{equation} \label{3.32}
u = \frac{x_{12}^2 x_{34}^2}{x_{13}^2 x_{24}^2}, \quad v = \frac{x_{12}^2 x_{34}^2}{x_{14}^2 x_{23}^2}
\end{equation}
The $\lambda_D$ are polynomials and the $\mu_D$ are rational functions
\begin{align} 
\lambda_1(u,v) &= 20-20Y+ \frac{10}{3}Y^2-\frac{8}{3}v(2-Y)+\frac{1}{3}v^2 \label{3.33} \\ 
\lambda_{15}(u,v) &= 2Y(2-Y)-vY \label{3.34} \\
\lambda_{20}(u,v) &= \frac{5}{3}Y^2 +\frac{5}{6}v(Y-2)+\frac{1}{6}v^2 \label{3.35} \\
\lambda_{84}(u,v) &= \frac{3}{2}v(2-Y) - \frac{1}{2}v^2 \label{3.36} \\
\lambda_{105}(u,v) &= v^2 \label{3.37} \\
\lambda_{175}(u,v) &= vY \label{3.38} 
\end{align}
using the shorthand
\begin{equation} \label{3.39}
Y= 1-\frac{v}{u}
\end{equation}
moreover
\begin{align}
\mu_1 &= \frac{4v}{1-Y}(\frac{20}{3}-\frac{10}{3}Y+\frac{1}{3}v) \label{3.40} \\
\mu_{15} &=\frac{4v}{1-Y} 2Y \label{3.41} \\
\mu_{20} &=\frac{4v}{1-Y} (\frac{10}{3}-\frac{5}{3}Y+\frac{1}{6}v)= \frac{1}{2}\mu_1 \label{3.42} \\
\mu_{84} &=\frac{4v}{1-Y} (- \frac{1}{2}v) = - \frac{1}{2}\mu_{105} \label{3.43} \\ 
\mu_{105} &=\frac{4v}{1-Y}v \label{3.44} \\ 
\mu_{175} &=0 \label{3.45} 
\end{align}
Finally we obtain for $\phi$ the two-variable hypergeometric series
\begin{equation} \label{3.46}
\phi(u,v) = -4v^2 \frac{\partial}{\partial \epsilon} \sum_{m,n=0}^{\infty} \frac{Y^m v^{n+\epsilon}}{m! \Gamma(n+1+\epsilon)} \frac{\Gamma(n+3+ \epsilon) \Gamma(n+m+3+ \epsilon) \Gamma(n+m+4+ \epsilon)}{\Gamma(2n+m+6+2 \epsilon)}|_{\epsilon=0}
\end{equation}

In the subsequent sections we will use this expression (\ref{3.31}) to derive four-point functions for multitrace operators at order $\mathcal{O}(\frac{1}{N^2})$.

\section{The $(2r+2)$-point functions at order $\mathcal{O}(1)$}
\setcounter{equation}{0}
We study the Green function
\begin{equation} \label{4.1}
\langle :\prod_{a=1}^r O_2^{I_a}(x_1^{(a)}): O_2^{I_{r+1}}(x_2):\prod_{b=1}^r O_2^{J_b}(x_3^{(b)}): O_2^{J_{r+1}}(x_4) \rangle
\end{equation}
at leading order $\mathcal{O}(\frac{1}{N^0})$ in an AdS calculation and take the limit
\begin{align} 
x_1^{(a)} &\longrightarrow x_1 \quad (\text{all}\; a) \label{4.2} \\
x_3^{(b)} &\longrightarrow x_3 \quad (\text{all}\; b) \label{4.3}
\end{align}
This limit is exact if we project the $SO(20)$ tensor of rank $r$ onto the $SO(6)$ representation $Y^{(6)}_{(2r)}$, since, due to Skiba's theorem, no renormalization is necessary. At leading order we can express (\ref{4.1}) by two projectors on irreducible $SO(20)$ representations defined by the Young tableaus\footnote{The latter are labelled by their partition.}
\begin{figure}[htb]
\begin{centering}
\includegraphics[scale=0.6]{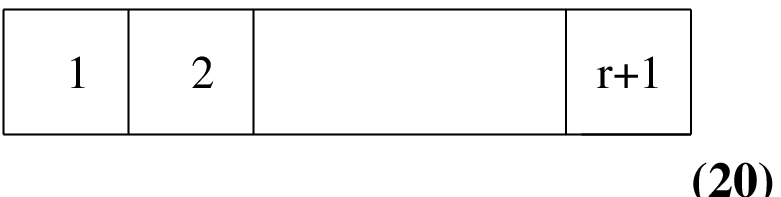}, \qquad \includegraphics[scale=0.6]{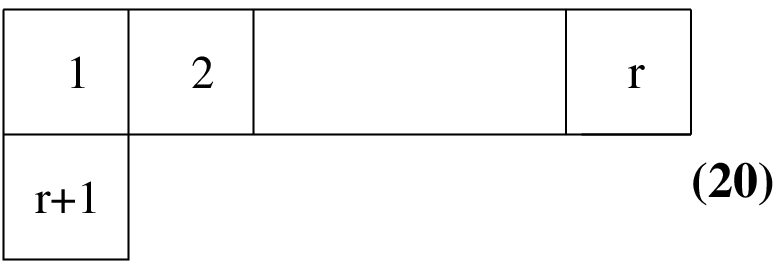} 
\caption{Representations $Y^{20}_{(r+1)}$ and $Y^{20}_{(r,1)}$ }
\end{centering}
\end{figure}

\noindent
which can be decomposed into irreducible $SO(6)$ representations containing the two sets of representations $Y^{(6)}_{(2r+2-2n, 2n)}$ and $Y^{(6)}_{(2r+1-2n, 2n+1)}$ at least, with $n \in \{0, 1, ..., [\frac{r+1}{2}]\}$ and $n \in \{0, 1, ..., [\frac{r}{2}]\}$, respectively.

According to the AdS rules, we get for (\ref{4.1}) in the limits (\ref{4.2}), (\ref{4.3})
\begin{equation} \label{4.4}
(x^2_{13})^{-2r}(x^2_{24})^{-2} \{(\sum_{s \in S_r}\prod_{a=1}^r \delta^{I_a J_{s(a)}}) \delta^{I_{r+1} J_{r+1}} + (\frac{v}{u})^2 \sum_{c,d=1}^r  \delta^{I_{r+1} J_{d}}  \delta^{I_{c} J_{r+1}}(\sum_{s \in S_{r-1}}\prod_{\substack{a=1 \\ a \neq c, s(a) \neq d}}^r \delta^{I_a J_{s(a)}})\}
\end{equation}
with
\begin{equation} \label{4.5}
\frac{v}{u} = \frac{x_{13}^2 x_{24}^2}{x_{14}^2 x_{23}^2} = 1-Y
\end{equation}
From the discussion in section 2 we know that\footnote{The normal product is elementary at order $\mathcal{O}(1)$.}
\begin{equation} \label{4.6}
:\prod_{\substack{a=1 \\ a \neq c}}^r O_2^{J_a}(x_4)[O_2^{J_c}(x_4) \partial_\nu O_2^{J_{r+1}}(x_4)- O_2^{J_{r+1}}(x_4)\partial_\nu O_2^{J_c}(x_4)]: = W_\nu^{J_1 J_2...(J_c)...J_r J_{r+1}}(x_4)
\end{equation}
is quasi-primary and has as field with transposed symmetry
\begin{equation} \label{4.7}
W_\mu^{T, I_1 I_2...I_r I_{r+1}}(x_2) = \frac{1}{r} \sum_{c=1}^r: \prod_{\substack{a=1 \\ (a \neq c)}}^r O_2^{I_a}(x_2)[O_2^{I_c}(x_2) \partial_\mu O_2^{I_{r+1}}(x_2)- O_2^{I_{r+1}}(x_2)\partial_\mu O_2^{I_c}(x_2)]:
\end{equation}
Then
\begin{multline} \label{4.8}
r \langle W_{\mu}^{T, I_1 I_2...I_r I_{r+1}}(x_2) W_{\nu}^{J_1 J_2...(J_c)...J_r J_{r+1}}(x_4) \rangle = 4(r+1)(\delta_{\mu\nu}-2\frac{x_{{24},\mu}x_{{24},\nu}}{x^2_{24}}) \\
\times [\delta^{I_{r+1}, J_{r+1}} (\sum_{s \in S_r} \prod_{a=1}^r \delta^{I_a J_{s(a)}})-\delta^{I_{r+1}, J_{c}} \sum_{b=1}^r \delta^{I_{b} J_{r+1}}(\sum_{s \in S_{r-1}} \prod_{\substack{a=1 \\ a \neq b, s(a) \neq c}}^r \delta^{I_a J_{s(a)}})]
\end{multline}
The square bracket defines the projector $P^{(20)}_{Y_{(r,1)}}$ with numbering as in Figure 9
\begin{figure}[htb]
\begin{centering}
\includegraphics[scale=0.6]{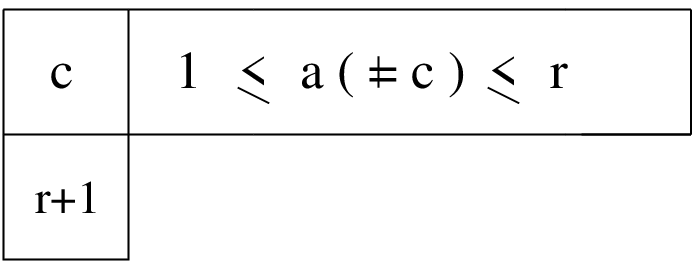} 
\caption{Representation $Y_{(r,1)}$ }
\end{centering}
\end{figure}

\noindent
This projector is automatically symmetric in $\{J_a\}^r_{\substack{a=1}{(a \neq c)}}$, so that the order of the $\{a\}$ in Figure 9 is irrelevant. Correctly normalized (see (\ref{2.9})), we get
\begin{equation} \label{4.9}
P^{(20)}_{Y^{(c)}_{(r,1)}} = \frac{1}{(r+1)(r-1)!} [...]_{({4.8})}
\end{equation}
If we sum over $c$
\begin{equation} \label{4.10}
\sum_c W_\nu^{J_1 J_2...(J_c)...J_r J_{r+1} }(x_4) = W_\nu^{J_1 J_2 ...J_r J_{r+1}}(x_4)
\end{equation}
and multiply $W_\mu^T$ by $r$, so that
\begin{equation} \label{4.11}
r W_\mu^{ T I_1 I_2...I_r I_{r+1}} = W_\mu^{I_1 I_2...I_r I_{r+1}}
\end{equation}
we obtain
\begin{multline} \label{4.12}
\langle W_\mu^{I_1 I_2...I_r I_{r+1}}(x_2)  W_\nu^{J_1 J_2...J_r J_{r+1}}(x_4) \rangle =  4(r+1)(\delta_{\mu\nu}-2\frac{x_{{24},\mu}x_{{24},\nu}}{x^2_{24}})(x_{24}^2)^{-2(r+1)-1} \\
[r \delta^{I_{r+1}, J_{r+1}} (\sum_{s \in S_r} \prod_{a=1}^r \delta^{I_a J_{s(a)}})- \sum_{b, c =1}^r \delta^{I_{r+1}, J_{c}} \delta^{I_{b}, J_{r+1}}(\sum_{s \in S_{r-1}} \prod_{\substack{a=1 \\ a \neq b, s(a) \neq c}}^r \delta^{I_a J_{s(a)}})]
\end{multline}
The bracket leads to the projector
\begin{equation} \label{4.13}
P^{(20)}_{Y^{(\text{sym})}_{(r,1)}} = \frac{1}{r} \sum_{c=1}^r P^{(20)}_{Y^{(c)}_{(r,1)}} = \frac{1}{(r+1)!} [...]_{({4.12})}
\end{equation}
The l.h.s. labels $\{I_a\}^r_{a=1}$ are symmetric in $P^{(20)}_{Y^{(c)}_{(r+1)}}$, so that for arbitrary $d$ 
\begin{equation} \label{4.14}
P^{(20)}_{Y^{(c)}_{(r,1)}} \cdot P^{(20)}_{Y^{(d)}_{(r,1)}} = P^{(20)}_{Y^{(d)}_{(r,1)}}
\end{equation}
since the asymmetric $(c)$ is averaged out.
It follows
\begin{equation} \label{4.15}
P^{(20)}_{Y^{(\text{sym})}_{(r,1)}} \cdot P^{(20)}_{Y^{(c)}_{(r,1)}} = P^{(20)}_{Y^{(c)}_{(r,1)}}
\end{equation}
and 
\begin{equation} \label{4.16}
 P^{(20)}_{Y^{(\mathrm{sym})}_{(r,1)}} \cdot  P^{(20)}_{Y^{(\mathrm{sym})}_{(r,1)}} = P^{(20)}_{Y^{(\mathrm{sym})}_{(r,1)}}
\end{equation}
as required.

Next we use the fact that at $x_{12} \rightarrow 0$, $x_{34}
\rightarrow 0$ we have (see (\ref{4.5}))
\begin{equation} \label{4.17}
Y = 2[(x_{12} x_{34}) - 2\frac{(x_{12}x_{24}) (x_{34}x_{24})}{x_{24}^2}] \frac{1}{x_{24}^2}(1+\mathcal{O}(x_{12}, x_{34}))
\end{equation}
so that the two-point function of $W_\mu$ (\ref{4.12}), at $x_{12} \rightarrow 0$, $x_{34} \rightarrow 0$ is
\begin{equation} \label{4.18}
x_{12,\mu} x_{34,\nu} \langle W_\mu^{I_1 I_2...I_r I_{r+1}}(x_2)  W_\nu^{J_1 J_2...J_r J_{r+1}}(x_4) \rangle  \cong (r+1)(r+1)!(1-(\frac{v}{u})^2)P^{I_1...I_{r+1},J_1...J_{r+1}}_{Y^{(\text{sym})}_{(r,1)}}  (x_{24}^2)^{-2(r+1)}
\end{equation}
Correspondingly the complete $(2r+2)$-point Green function at order $\mathcal{O}(1)$ is decomposed into irreducible representations of $SO(20)$ as
\begin{equation} \label{4.19}
(x_{13}^2)^{-2r} (x_{24}^2)^{-2} (A P_{Y_{(r+1)}} + B P_{Y^{(\mathrm{sym})}_{(r,1)}})
\end{equation}
with
\begin{align}
A &= r! (1+r(\frac{v}{u})^2) \label{4.20} \\
B &= r! (1-(\frac{v}{u})^2) \label{4.21}
\end{align}
Thus we obtain one tower of quasi-primary fields for each $SO(20)$ irreducible representation $Y_{(r+1)}^{(20)}$ and $Y_{(r,1)}^{(20)}$ with the lowest rank tensor fields $:\prod_{a=1}^{r+1} O_{2}^{I_a}(x):$ and $W_{\mu}^{I_1 ... I_{r+1}}(x)$.

In view of the analysis in section $6$, we decompose these $\bf{R}_{20}$ projectors into $\bf{R}_{6}$ projectors (with pairings) according to
\begin{equation} \label{4.22}
P^{(20)}_{Y_{(r+1)}} = \sum_{n=0}^{[\frac{r+1}{2}]} P^{(6)}_{Y_{(2r+2-2n,2n)}} + 3 \, \text{row irreps} \, + \, \text{contraction irreps}
\end{equation}
and 
\begin{equation} \label{4.23}
P^{(20)}_{Y_{(r,1)}} = \sum_{n=0}^{[\frac{r}{2}]} P^{(6)}_{Y_{(2r+1-2n,2n+1)}} + 3 \, \text{row irreps} \, + \, \text{contraction irreps}
\end{equation}


\section{The $(2r+2)$-point functions at order $\mathcal{O}(\frac{1}{N^2})$}
\setcounter{equation}{0}

Perturbative corrections to the Green function (\ref{4.1}) at order $\mathcal{O}(\frac{1}{N^2})$ are obtained by insertion of a pair of $3$-vertex graphs or from the complete sum over $4$-vertex graphs\footnote{Each graph contained in this sum is connected.} which we denote by
\begin{figure}[htb]
\begin{centering}
\includegraphics[scale=0.5]{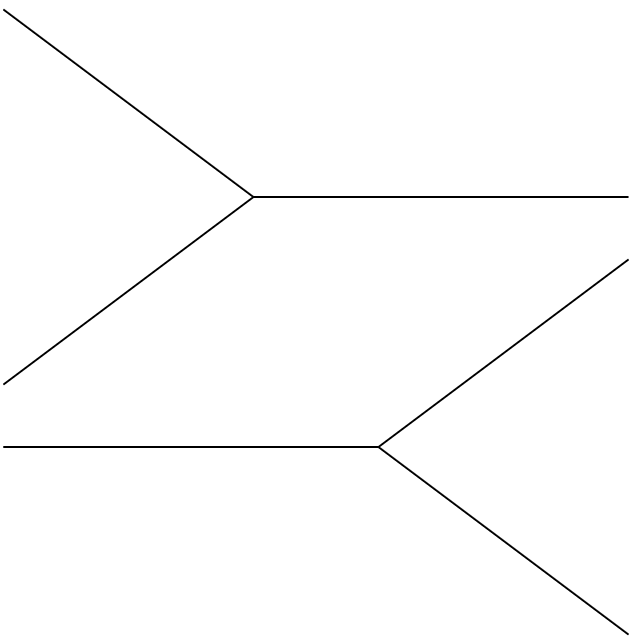} \qquad \includegraphics[scale=0.85]{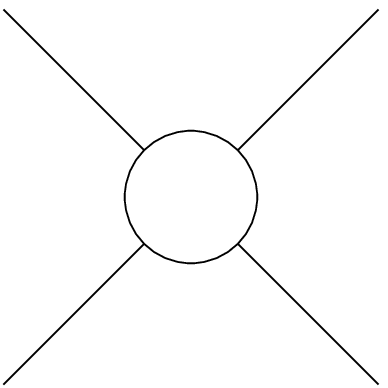}
\caption{Pair of $3$-vertex graphs and dressed $4$-vertex graph}
\end{centering}
\end{figure}

\noindent
A single $3$-vertex graph of $O^I_2$-fields
\begin{figure}[htb]
\begin{centering}
\includegraphics[scale=0.6]{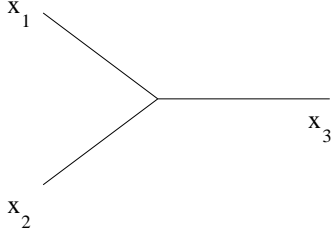}
\caption{Single $3$-vertex graph with coordinates}
\end{centering}
\end{figure}

\noindent
gives a contribution
\begin{equation} \label{5.1}
\frac{8^{\frac{1}{2}}}{N} \frac{C^{I_1 I_2 I_3}} {x_{12}^2 x_{23}^2 x_{31}^2} (1+\mathcal{O}(\frac{1}{N^2}))
\end{equation}
Going over to $SO(6)$ tensor notation by 
\begin{equation} \label{5.2}
C^{I_1 I_2 J} \longrightarrow \sum_{I_{1,2},J} C_{i_1 i_2}^{I_1} C_{i_3 i_4}^{I_2} C_{j_1 j_2}^{J} C^{I_1 I_2 J}
\end{equation}
we obtain
\begin{align} \label{5.3}
\frac{1}{4} \{ \delta_{i_1 i_3} P_{20}^{i_2 i_4, j_1 j_2} +  \delta_{i_1 i_4} P_{20}^{i_2 i_3, j_1 j_2} + \delta_{i_2 i_3} P_{20}^{i_1 i_4, j_1 j_2} &+ \delta_{i_2 i_4} P_{20}^{i_1 i_3, j_1 j_2}\} \notag \\
&- \frac{1}{6} \{\delta_{i_1 i_2} P_{20}^{i_3 i_4, j_1 j_2} + \delta_{i_3 i_4} P_{20}^{i_1 i_2, j_1 j_2}\} 
\end{align}
where $ P_{20}^{(6)}$ was given on the r.h.s. of (\ref{3.4}). This expression contains one $SO(6)$ contraction $\delta_{i_a i_b}$ in each term but does not appear as ''trace term'' of another $SO(6)$ tensor. Therefore in the direct channel of the $(2r+2)$-point function it yields  $SO(6)$ representations with $2r'+2$ blocks where
\begin{equation} \label{5.4}
r' \leq r-1
\end{equation}
However, we select $SO(6)$ representations which conserve the number of blocks in each step of fusion of single-trace to multiple-trace operators. Then the $3$-vertex graphs can be neglected at $\mathcal{O}(\frac{1}{N^2})$. Singular terms ($(x_{12}^2)^{-1}$ in (\ref{5.1})) in the operator product expansion are completely eliminated in this fashion.
From the $4$-vertex insertion only the projections on $Y_{(4)}$, $Y_{(3,1)}$, $Y_{(2,2)}$ are needed, which in the limit $v \rightarrow 0$, $Y \rightarrow 0$, corresponding to $x_{12} \rightarrow 0$,$x_{34} \rightarrow 0$ give the corrections
\begin{align} 
 &Y_{(4)}: \qquad \sigma_{105} = +4 \qquad (\text{dim} \, = 105), \label{5.5} \\
 &Y_{(2,2)}: \qquad \sigma_{84}= -2  \qquad (\text{dim} \, = 84), \label{5.6} \\
 &Y_{(3,1)}: \qquad \sigma_{175}= 0  \qquad (\text{dim} \, = 175), \label{5.7} 
\end{align}
where $\sigma_D$ is defined by
\begin{equation*}
\sigma_D=\frac{1}{v^2}(\lambda_D \phi +\mu_D)|_{v=Y=0}.
\end{equation*}
and the corrections are given in terms of $\frac{\sigma}{N^2}$.

\noindent
The result (\ref{5.7}) is trivial, since a vector field must arise in the limit $x_{12} \rightarrow 0$,$x_{34} \rightarrow 0$, which is indeed contained (through one power of $Y$, see (\ref{2.18}), (\ref{4.17}))
\begin{equation} \label{5.8}
Y_{(3,1)}: \quad \tfrac{1}{v^2}(\lambda_{175} \phi +\mu_{175}) \cong - \tfrac{4}{5} Y v(log \,v+\tfrac{4}{3}) + \; \text{higher order corrections.} \,.
\end{equation}
due to (\ref{4.18}).

Now we turn to the $\mathcal{O}(\frac{1}{N^2})$ correction of the normal product
\begin{equation} \label{5.9}
\underset{\substack{x^{(a)} \rightarrow x \\(\text{all} \; a)}}{lim} :\prod_{a=1}^r O_2^{(I_a)}(x^a):
\end{equation}
which at leading order could be projected onto one $SO(20)$ representation but at order $\mathcal{O}(\frac{1}{N^2})$ must be projected on independent irreducible $SO(6)$ representations. Block number conservation allows only the representations of two rows
\begin{equation} \label{5.10}
Y^{(6)}_{(2r-n,n)}, \qquad n \in \{0, 1, 2,...\}
\end{equation}
and three row representations which must further be projected on self-dual and anti-self-dual representations. In the case of the normal product (\ref{5.9}) we select only the representation $Y^{(6)}_{(2r)}$, whereas in the analysis of the $(2r+2)$-point function (\ref{4.1}) we allow all representations
\begin{equation} \label{5.11}
Y^{(6)}_{(2r+2-n,n)}, \quad n \in \{0,1,2,..\}
\end{equation}
The totally symmetric case in (\ref{5.10}), (\ref{5.11}), i.e. $n=0$, is simple. Due to Skiba's theorem, it can be computed by free field theory\footnote{In this case the limit (\ref{5.5}) exists trivially.}
\begin{equation} \label{5.12}
\Omega_{i_1 i_2 ...i_{2r}}(x) = \sum_{\{I\}} Q_{Y_{2r i_1 i_2 ...i_{2r}}}^{I_1 I_2 ...I_r} :\prod_{a=1}^r O_2^{I_a}:
\end{equation}
and 
\begin{equation} \label{5.13}
\langle \Omega_{i_1 i_2 ...i_{2r}}(x)\Omega_{j_1 j_2 ...j_{2r}}(y) \rangle = P_{Y_{(2r)}}^{i_1 i_2 ...i_{2r},j_1 j_2 ...j_{2r}} r! (1+\frac{r(r-1)}{N^2}+\mathcal{O}(\frac{1}{N^4}))((x-y)^2)^{-2r}
\end{equation}

This result can be proven easily from (\ref{5.5}) and simple combinatorics of embedding the $4$-vertex in all possible ways (see Figure 12).
\begin{figure}[htb!]
\begin{centering}
\includegraphics[scale=0.45]{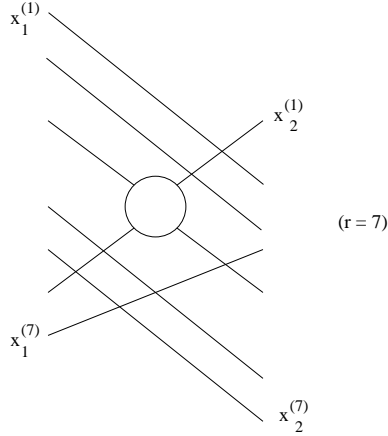}
\caption{Graphs for the $2$-point function of the normal product $\Omega$}
\end{centering}
\end{figure}

\noindent
We thus obtain (see (\ref{5.5}))
\begin{equation} \label{5.14}
{\binom{r}{2}}^2 (r-2)! \frac{4}{N^2} = r! \frac{r(r-1)}{N^2}
\end{equation}
which for $r=2$ gives back (\ref{5.5}).

For the $(2r+2)$-point function this yields four classes of contributions at $\mathcal{O}(\frac{1}{N^2})$. First we have the class $(0)$ of  disconnected graphs with the propagator product
\begin{equation} \label{5.15}
\frac{\delta^{I_{r+1} J_{r+1}}}{(x_{24}^2)^2} \times \text{two-point function} \;(\ref{5.13})
\end{equation}
Inserting (\ref{5.13}) we obtain (projectors on irreducible representations of $SO(6)$ everywhere from now on)
\begin{equation} \label{5.16}
(x_{13}^2)^{-2r} (x_{24}^2)^{-2} \frac{r(r-1)}{N^2} r! P_{Y_{(2r)}}^{I_1 I_2 ...I_{r},J_1 J_2 ...J_{r}} \delta^{I_{r+1} J_{r+1}}
\end{equation}
There are four additional classes of connected graphs, the first three of which have one pair of legs of the $4$-point insertion at the same coordinate $x_1$ (or $x_3$ or both) (see Figure 13).

\begin{figure}[htb]
\begin{centering}
\includegraphics[scale=0.45]{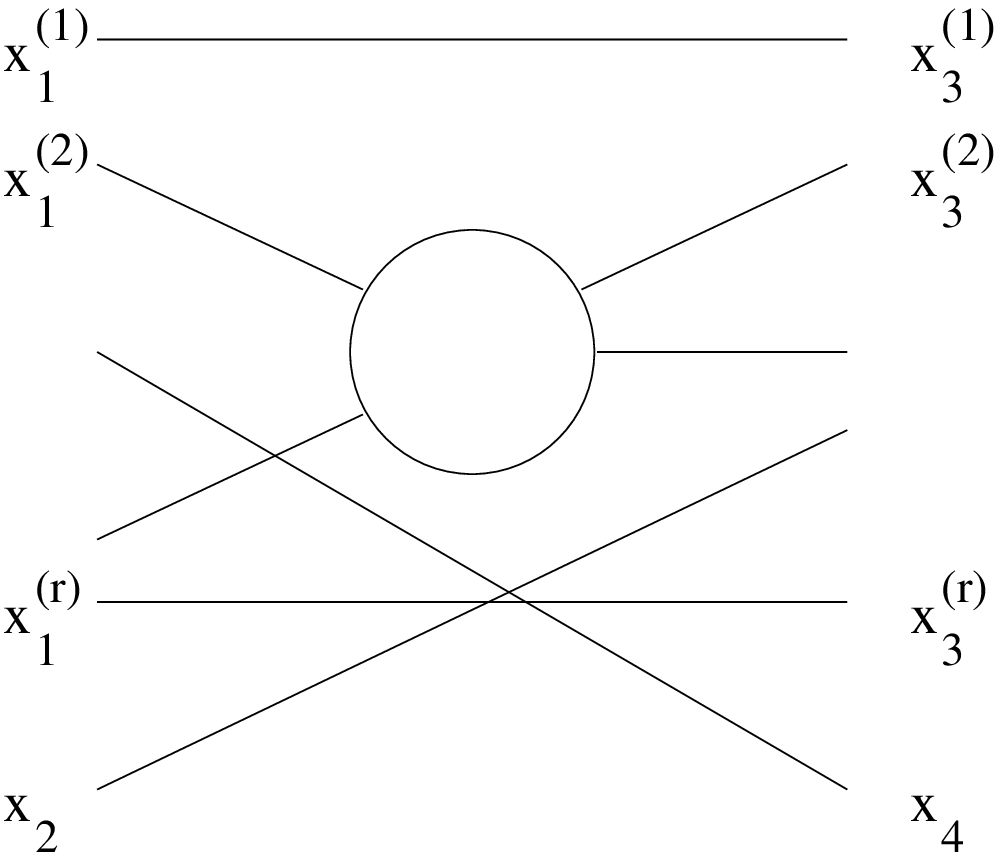} \qquad \includegraphics[scale=0.45]{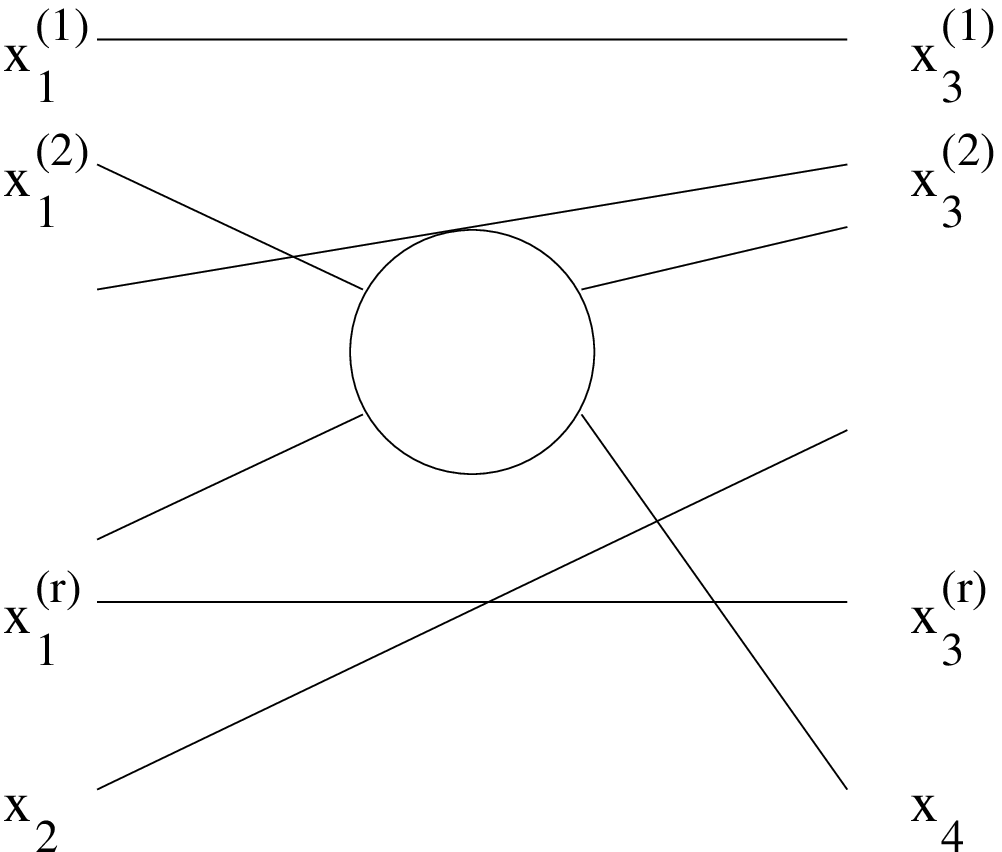}
\caption{Class $(1)$ graphs with both pairs of legs and class $(2)$ graphs with the left pair of legs of the $4$-vertex insertion coinciding }
\end{centering}
\end{figure}

\noindent
The graphs of the third class are obtained from the graphs of the second class by forward-backward reflection, while the fourth class of graphs makes full use of the $4$-point insertion (see Figure 14).
\begin{figure}[htb]
\begin{centering}
\includegraphics[scale=0.45]{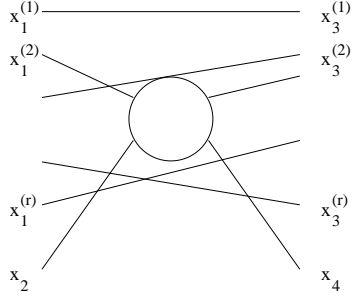}
\caption{Class $(4)$ graphs}
\end{centering}
\end{figure}

\noindent
In calculating the contributions of the graphs, we single out those terms for which
\begin{enumerate}
\item
The $SO(6)$ representation from $\{x_1^{(a)}\}$ and  $\{x_3^{(b)}\}$ is projected on $Y_{(2r)}^{(6)}$.
\item 
The total number of blocks is conserved ($r+1$ for $SO(20)$, $2r+2$ for $SO(6)$)
\item
To make sense $r$ must be bigger or equal $2$ in the rest of this section.
\end{enumerate}
In the case of class $(1)$ graphs, we obtain (before the projection on $Y_{(2r)}^{(6)}$)
\begin{equation} \label{5.17}
(x_{13}^2)^{-2r} (x_{24}^2)^{-2} (\frac{v}{u})^2 \sum_{D \in \{{\bf{84}}, {\bf{105}}, {\bf{175}}\}} \negthickspace \sum_{\substack{a,b,c;\\ s \in S_r \\ (a > b, s(a)>s(b))}} \negthickspace \negthickspace \negthickspace  \negthickspace \negthickspace \negthickspace \negthickspace
\frac{\sigma_D}{N^2} P_D^{I_a I_b, J_{s(a)} J_{s(b)}} \delta^{I_c J_{r+1}} \delta^{I_{r+1} J_{s(c)}} \negthickspace \negthickspace \prod^r_{\substack{i=1 \\ i \notin \{a,b,c\}}} \negthickspace \negthickspace \delta^{I_i J_{s(i)}}
\end{equation}
In the case of class $(2)$ graphs, we obtain
\begin{equation} \label{5.18}
(x_{13}^2)^{-2r} (x_{24}^2)^{-2}(\frac{v}{u})^2 \sum_{D \in \{{\bf{84}}, {\bf{105}}, {\bf{175}}\}} \sum_{\substack{a,b;\\ s \in S_r \\ (a > b)}} \frac{\sigma_D}{N^2} P_D^{I_a I_b, J_{s(a)} J_{r+1}}  \delta^{I_{r+1} J_{s(b)}} \prod^r_{\substack{i=1 \\ i \notin \{a,b\}}} \delta^{I_i J_{s(i)}}
\end{equation}
and for case class $(3)$ graphs, which are obtained from class $(2)$ graphs by a left-right reflection in the same fashion
\begin{equation} \label{5.19}
(x_{13}^2)^{-2r} (x_{24}^2)^{-2}(\frac{v}{u})^2 \sum_{D \in \{{\bf{84}}, {\bf{105}}, {\bf{175}}\}} \sum_{\substack{a,b;\\ s \in S_r \\ (s(a) >s(b))}}  \frac{\sigma_D}{N^2} P_D^{I_a I_{r+1}, J_{s(a)} J_{s(b)}} \delta^{I_{b} J_{r+1}} \prod^r_{\substack{i=1 \\ i \notin  \{a,b\}}} \delta^{I_i J_{s(i)}}
\end{equation}

These classes (1), (2), (3) have a common property: If we project on $Y_{(2r)}^{(6)}$ only the representation $Y_{(4)}^{(6)}$ ($D=\bf{105}$) in the sum over $D$ survives.

Now we come to class (4) which yields
\begin{equation} \label{5.20} 
(x_{13}^2)^{-2r} (x_{24}^2)^{-2} u^{-2} \negthickspace \negthickspace \sum_{D \in \{84, 105, 175\}} \sum_{\substack{a \\ s \in S_r}}  \frac{1}{N^2}  P_D^{I_a I_{r+1}, J_{s(a)} J_{r+1}} (\lambda_D \phi + \mu_D)(u,v) \; \prod_{\substack{i=1 \\ \neq a}}^r  \delta^{I_i J_{s(i)}}
\end{equation}
Notice that the factor in front of the sum is $u^{-2}$ instead of $(\frac{u}{v})^{-2}$ since in $\lambda_D \phi + \mu_D$ the factor $v^2$ has not been extracted.

Next we project first on $Y_{(2r)}^{(6)}$ with labels $\{I_a, J_a\}_{a=1}^r$, then on  $Y_{(2r+2-n, n)}^{(6)}$, $n \in \{0,1,2\}$ with labels  $\{I_a, J_a\}_{a=1}^{r+1}$. 
The two projectors can always be contracted into one, since
\begin{equation} \label{5.21}
Y_{(2r+2-n, n)}^{(6)} Y_{(2r)}^{(6)} = Y_{(2r+2-n, n)}^{(6)}, \qquad n \in \{0,1,2\}
\end{equation}
We start with (\ref{5.16}). In this case the first projection is trivial. We expand then
\begin{equation} \label{5.22}
 P_{Y_{(2r)}}^{I_1...I_{r}, J_1...J_{r}} \delta^{I_{r+1} J_{r+1}} = \sum_{n=0}^{r+1} A_{r,n} P_{Y_{(2r+2-n, n)}}^{I_1...I_{r}I_{r+1}, J_1...J_{r}J_{r+1}} +\text{contraction terms}
\end{equation}
On the l.h.s. we have a tensor product of $Y_{(2r)}^{(6)}$ with $Y_{(2)}^{(6)}$, implying obviously (see Figure 15)
\begin{equation} \label{5.23}
A_{r,0} = A_{r,1} = A_{r,2} = 1; \quad A_{r,n} =0, \; n > 2
\end{equation}
\begin{figure}[htb]
\begin{centering}
\includegraphics[scale=0.6]{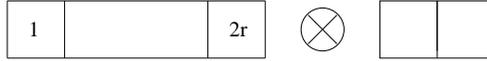}
\caption{Product of $Y_{(2r)}^{(6)}$ with $Y_{(2)}^{(6)}$}
\end{centering}
\end{figure}

\noindent
All graphs are in standard symmetrization.

In the case of (\ref{5.17}), we obtain similarly
\begin{equation} \label{5.24}
\frac{1}{N^2} (\frac{v}{u})^2 r! r(r-1)(r-2) P_{Y_{(2r+2)}}
\end{equation}
In fact, projecting on $P_{Y_{(2r+2)}}$ yields
\begin{multline} 
P_{Y_{(2r+2)}}^{j_1 j_2...j_{2r} j_{2r+1} j_{2r+2}, i_1 i_2...i_{2r} i_{2r+1} i_{2r+2}}  P_{Y_{(4)}}^{i_1...i_{4}, j_1...j_{4}} \\ P_{Y_{(2)}}^{i_5 i_6, j_{2r+1} j_{2r+2}}   P_{Y_{(2)}}^{i_ {2r+1} i_{2r+2}, j_{5} j_{6}} \prod _{a=4}^r  P_{Y_{(2)}}^{i_{2a-1} i_{2a}, j_{2a-1} j_{2a}} \\
= P_{Y_{(2r+2)}}^{i_1 i_2 i_3 i_4 i_{2r+1} i_{2r+2} i_5 i_6...i_{2r}, i_1 i_2...i_{2r} i_{2r+1} i_{2r+2}} = \,\text{dim} \,(P_{Y_{(2r+2)}}) \label{5.25}
\end{multline}
since total symmetry of both the r.h.s. and the l.h.s. allows the contraction of all indices\footnote{The underlying ''contraction theorems'' are discussed in Appendix A.}.
For the projection on $ P_{Y_{(2r,2)}}$ the procedure is very similar, namely
\begin{multline}
P_{Y_{(2r,2)}}^{j_1 j_2...j_{2r} j_{2r+1} j_{2r+2}, i_1 i_2...i_{2r} i_{2r+1} i_{2r+2}}  P_{Y_{(4)}}^{i_1...i_{4}, j_1...j_{4}} P_{Y_{(2)}}^{i_5 i_6, j_{2r+1} j_{2r+2}}\\  P_{Y_{(2)}}^{i_ {2r+1} i_{2r+2}, j_{5} j_{6}}  \prod _{a=4}^r  P_{Y_{(2)}}^{i_{2a-1} i_{2a}, j_{2a-1} j_{2a}} \\ = P_{Y_{(2r,2)}}^{i_1 i_2 i_3 i_4 i_{2r+1} i_{2r+2}, i_7...i_{2r}, i_5 i_6, i_1 i_2...i_{2r} i_{2r+1} i_{2r+2}} \label{5.26}
\end{multline}
Now the r.h.s. is antisymmetric in $\{i_1, i_{2r+1}\}$ and the l.h.s. is symmetric in these labels, yielding zero as a result. Relying on the same argument, the projection on $Y_{(2r+1,1)}$ vanishes as well.

Turning to (\ref{5.18}), the result is given by
\begin{equation} \label{5.27}
\frac{2}{N^2} (\frac{v}{u})^2 r!\, r(r-1)[P_{Y_{(2r+2)}}+ \frac{1}{6} P_{Y_{(2r,2)}}]
\end{equation}

For $P_{Y_{2r+2}}$ the argument given above can be carried over and the result is obviously correct. For $P_{Y_{(2r,2)}}$ we obtain
\begin{multline} \label{5.28}
P_{Y_{(2r,2)}}^{j_1 j_2...j_{2r} j_{2r+1} j_{2r+2}, i_1 i_2...i_{2r} i_{2r+1} i_{2r+2}} P_{Y_{(4)}}^{i_1...i_{4}, j_1 j_2 j_{2r+1} j_{2r+2}} P_{Y_{(2)}}^{i_ {2r+1} i_{2r+2}, j_{3} j_{4}} \prod _{a=3}^r  P_{Y_{(2)}}^{i_ {2a-1} i_{2a}, j_{2a-1} j_{2a}} \\
= P_{Y_{(2r,2)}}^{j_1 j_2...j_{2r} j_{2r+1} j_{2r+2}, (j_1 j_2 j_{2r+1} j_{2r+2}) j_{5} j_{6}...j_{2r} j_{3} j_{4}}
\end{multline}
where the bracket denotes total symmetrization. 
Contraction with respect to the labels $\{j_5 j_6...j_{2r}\}$ yields 
\begin{multline} \label{5.29}
(\text{dim} Y_{(2r,2)}) \cdot (\text{dim} Y_{(4,2)})^{-1} \times P_{Y_{(4,2)}}^{j_1 j_2 j_3 j_4 j_{2r+1} j_{2r+2}, (j_1 j_2 j_{2r+1} j_{2r+2}) j_3 j_4} \\
= \frac{1}{6} P_{Y_{(4,2)}}^{j_1 j_2 j_3 j_4 j_{2r+1} j_{2r+2}, j_{2r+1} j_{2r+2} j_1 j_2 j_3 j_4},
\end{multline}
with standard symmetrization of pairs. This argument is based on the total symmetry of the l.h.s. in $(j_1, j_2, j_3, j_4)$. This symmetry and an antisymmetry of the r.h.s. in an appropriate pair $(j_a,j_b)$ gives zero. Consider for example the label set on the r.h.s.
\begin{equation*} 
j_1 j_5 j_6 j_2 j_3 j_4 \notag
\end{equation*} 
where $2r+1$ was replaced by $5$ and $2r+2$ by $6$.
The corresponding contribution in $P$ is obtained from the symmetrizer
\begin{equation} \label{5.30}
(e-(1,3))(e-(4,5)) \Sigma_4 (1,2,5,6) \Sigma_2 (3,4)
\end{equation}
where $\sum_n = \sum_{s \in S_n} s$.
Symmetry of the l.h.s. in $(j_1 j_3)$ amounts to left multiplication of (\ref{5.30}) with
\begin{equation*} 
\frac{1}{2} (e+(1,3)) 
\end{equation*}
which gives zero. In this fashion we obtain the result (\ref{5.29}), understood to be in standard pairing, i.e. consisting of four terms for $(j_{2r+1} j_{2r+2})(j_1 j_2)$. Then the factor results from
\begin{equation} \label{5.31}
\frac{1}{6}= 4 \cdot \frac{1}{24}
\end{equation}

Once again using the symmetry of the l.h.s., we write
\begin{align}
&=  \frac{1}{6} P_{Y_{(4,2)}}^{j_3 j_4 j_1 j_2 j_{2r+1} j_{2r+2}, j_{2r+1} j_{2r+2} j_1 j_2 j_3 j_4} \notag \\
&=  \frac{1}{6} P_{Y_{(2,2)}}^{j_3 j_4 j_{2r+1} j_{2r+2}, j_{2r+1} j_{2r+2} j_3 j_4} \times (\text{dim} Y_{(4,2)})(\text{dim} Y_{(2,2)})^{-1} \label{5.32}
\end{align}
For the symmetrizer $e(2,2)$ with numbering as is figure 16
\begin{figure}[htb]
\begin{centering}
\includegraphics[scale=0.75]{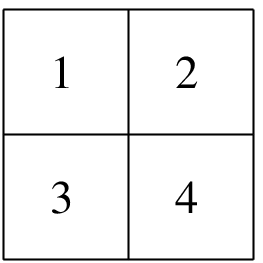}
\caption{Symmetrizer $e(2,2)$}
\end{centering}
\end{figure}

\noindent
we obtain
\begin{equation} \label{5.33}
(1,3)(2,4) e(2,2) = e(2,2)(1,3)(2,4) =  e(2,2)
\end{equation}
since
\begin{equation} \label{5.34}
(1,3)(2,4) (e-(1,3))(e-(2,4))  = (e-(1,3))(e-(2,4)(1,3)(2,4)) = (e-(1,3))(e-(2,4))
\end{equation}
and 
\begin{equation} \label{5.35}
(1,3)(2,4)(e+(1,2))(e+(3,4)) (1,3)(2,4) = (e+(1,2))(e+(3,4))
\end{equation}
Correspondingly we can continue (\ref{5.32})
\begin{equation} \label{5.36}
\frac{1}{6} P_{Y_{(2,2)}}^{j_3 j_4 j_{2r+1} j_{2r+2},  j_3 j_4 j_{2r+1} j_{2r+2}} = \frac{1}{6} \, \text{dim} \, Y_{(2,2)}
\end{equation}
A similar argument applies if we project onto $Y_{(2r+1,1)}$. Instead of (\ref{5.29}) we get in fact 
\begin{multline} \label{5.37}
P_{Y_{(5,1)}}^{j_1 j_2 j_3 j_4 j_{2r+1} j_{2r+2}; (j_1 j_2 j_{2r+1} j_{2r+2}) j_3 j_4} \\
= \frac{1}{4} P_{Y_{(5,1)}}^{j_1 j_2 j_3 j_4 j_{2r+1} j_{2r+2}, j_{2r+2} j_1 j_2 j_{2r+1} j_3 j_4} \\
= -\frac{1}{4} P_{Y_{(5,1)}}^{j_1 j_2 j_3 j_4 j_{2r+1} j_{2r+2}, j_4 j_1 j_2 j_{2r+1} j_3 j_{2r+2}}.
\end{multline}
since the l.h.s. labels can be reordered giving, as result, the r.h.s. labels. We thus get
\begin{equation} \label{5.38}
= - \frac{1}{4} \, \text{dim} \, Y_{(5,1)}
\end{equation} 
Finally we mention that from (\ref{5.19}) we get the same result as from (\ref{5.18})\footnote{The projectors used in the calculations are hermitean.}.

Now we consider the evaluation of (\ref{5.20}) and find as a result
\begin{equation} \label{5.39}
\frac{r!}{N^2} r u^{-2} [(\lambda_{\bf 84} \phi + \mu_{\bf 84}) P_{Y_{(2r,2)}} + (\lambda_{\bf 105} \phi + \mu_{\bf 105}) P_{Y_{(2r+2)}} +(\lambda_{\bf 175} \phi + \mu_{\bf 175}) P_{Y_{(2r+1,1)}}]
\end{equation}
For this result we have to evaluate the contractions
\begin{multline}   \label{5.40}
P_{Y_{(2r+2)}}^{j_1 j_2...j_{2r} j_{2r+1} j_{2r+2}, i_1 i_2...i_{2r} i_{2r+1} i_{2r+2}} P_{Y_{(4)}}^{i_1 i_{2} i_{2r+1} i_{2r+2}, j_1 j_2 j_{2r+1} j_{2r+2}} \prod _{a=2}^r  P_{Y_{(2)}}^{i_{2a-1} i_{2a}, j_{2a-1} j_{2a}} \\
= P_{Y_{(2r+2)}}^{j_1 j_2...j_{2r} j_{2r+1} j_{2r+2}, j_1 j_2...j_{2r} j_{2r+1} j_{2r+2}} = \text{dim}(Y_{(2r+2)})
\end{multline}
\begin{equation} \label{5.41}
P_{Y_{(2r+2)}}^{j_1 j_2...j_{2r} j_{2r+1} j_{2r+2}, i_1 i_2...i_{2r} i_{2r+1} i_{2r+2}} P_{Y_{(2,2)}}^{i_1 i_{2} i_{2r+1} i_{2r+2}, j_1 j_2 j_{2r+1} j_{2r+2}} \prod _{a=2}^r  P_{Y_{(2)}}^{i_{2a-1} i_{2a}, j_{2a-1} j_{2a}} = 0 
\end{equation}
\begin{equation} \label{5.42}
P_{Y_{(2r,2)}}^{j_1 j_2...j_{2r} j_{2r+1} j_{2r+2}, i_1 i_2...i_{2r} i_{2r+1} i_{2r+2}} P_{Y_{(4)}}^{i_1 i_{2} i_{2r+1} i_{2r+2}, j_1 j_2 j_{2r+1} j_{2r+2}} \prod _{a=2}^r  P_{Y_{(2)}}^{i_{2a-1} i_{2a}, j_{2a-1} j_{2a}} = 0 
\end{equation}
The representation $Y_{(2r,2)}$ would have appeared if we had changed the numbering to 
\begin{equation*} \label{5.43}
P_{Y_{(2r,2)}}^{j_1 j_2 j_{2r+1} j_{2r+2} j_3...j_{2r}, i_1 i_2 i_{2r+1} i_{2r+2} i_3...i_{2r}}
\end{equation*}
which represents a different fusion channel.
\begin{align} 
&P_{Y_{(2r,2)}}^{j_1 j_2...j_{2r} j_{2r+1} j_{2r+2}, i_1 i_2...i_{2r} i_{2r+1} i_{2r+2}}
\cdot  P_{Y_{(2,2)}}^{i_1 i_{2} i_{2r+1} i_{2r+2}, j_1 j_2 j_{2r+1} j_{2r+2}} \prod _{a=2}^r  P_{Y_{(2)}}^{i_{2a-1} i_{2a}, j_{2a-1} j_{2a}} \notag \\
&=  P_{Y_{(2,2)}}^{j_1 j_2 j_{2r+1} j_{2r+2},  i_1 i_2 i_{2r+1} i_{2r+2}}  P_{Y_{(2,2)}}^{i_1 i_2 i_{2r+1} i_{2r+2}, j_1 j_2 j_{2r+1} j_{2r+2}}
\cdot (\text{dim}Y_{(2r,2)}) (\text{dim}Y_{(2,2)})^{-1} = \text{dim}Y_{(2r,2)} \label{5.44}
\end{align}
For the projection on $P_{Y_{(2r+1,1)}}$ we can repeat the arguments of (\ref{5.42}), (\ref{5.44})
\begin{align}
&\text{Tr}(P_{Y_{(2r+1,1)}} P_{Y_{(4)}} \otimes \; \bigotimes_a  P_{Y_{(2)}}^{(a)}) = 0 \label{5.45} \\
&\text{Tr}(P_{Y_{(2r+1,1)}} P_{Y_{(2,2)}} \otimes \; \bigotimes_a  P_{Y_{(2)}}^{(a)}) = 0 \label{5.46} \\
&\text{Tr}(P_{Y_{(2r+1,1)}} P_{Y_{(3,1)}} \otimes \; \bigotimes_a  P_{Y_{(2)}}^{(a)}) = \text{dim} \, Y_{(2r+1,1)} \label{5.47} 
\end{align}

Summing up the results for (\ref{5.16})-(\ref{5.20}) (see Table 1) , we get for the order $\mathcal{O}(\frac{1}{N^2})$
\begin{multline} \label{5.48}
+ \frac{r!}{N^2} \{ r(r-1)[(1+(r+2)(\frac{v}{u})^2)) P_{Y_{(2r+2)}} + (1+ \frac{2}{3} (\frac{v}{u})^2) P_{Y_{(2r,2)}} \\
+ (1-(\frac{v}{u})^2) P_{Y_{(2r+1,1)}}] + r u^{-2} \sum_{n \in \{0,1,2\}} (\lambda_{4-n,n} \phi + \mu_{4-n,n})  P_{Y_{(2r+2-n,n)}} \}
\end{multline}
\begin{table}[htb]
\begin{centering}
\begin{tabular}{|c|c|c|c|c|}
        \hline
	eqn.&factor&$ P_{Y_{(2r+2)}}$&$ P_{Y_{(2r,2)}}$&$P_{Y_{(2r+1,1)}}$\\
        \hline
        (\ref{5.16})&$\frac{r!}{N^2}$&$r(r-1)$&$r(r-1)$&$r(r-1)$ \\ \hline
        (\ref{5.17})&$\frac{r!}{N^2}(\frac{v}{u})^2$&$r(r-1)(r-2)$&$0$&$0$ \\ \hline 
        (\ref{5.18})&$\frac{r!}{N^2}(\frac{v}{u})^2$&$2r(r-1)$&$\frac{1}{3}r(r-1)$&$-\frac{1}{2}r(r-1)$\\ \hline
        (\ref{5.19})&$\frac{r!}{N^2}(\frac{v}{u})^2$&$2r(r-1)$&$\frac{1}{3}r(r-1)$&$-\frac{1}{2}r(r-1)$\\ \hline
        (\ref{5.20})&$\frac{r!}{N^2} u^{-2}$&$r(\lambda_{105} \phi +\mu_{105})$&$r(\lambda_{84} \phi +\mu_{84})$&$r(\lambda_{175} \phi +\mu_{175})$\\ 
        \hline
\end{tabular} 
\caption{Table of $\frac{1}{N^2}$ contributions to the $4$-point function (\ref{4.1}) for each class of graphs}
\end{centering}
\end{table}
Adding the $\mathcal{O}(1)$ terms (\ref{4.20})-(\ref{4.22}) and decomposing $P^{(20)}_{Y_{(2r+1)}}$ into $SO(6)$ representation projectors as in (\ref{4.23}) yields
\begin{multline} 
(x_{13}^2)^{-2r} (x_{24}^2)^{-2} r! \{[(1+r(\frac{v}{u})^2)+\frac{1}{N^2}[r(r-1)(1+(r+2)(\frac{v}{u})^2)+ r u^{-2} (\lambda_{105} \phi + \mu_{105})] P_{Y_{(2r+2)}}  \\
+[(1+r(\frac{v}{u})^2)+ \frac{1}{N^2}[r(r-1)(1+ \frac{2}{3}(\frac{v}{u})^2) +r u^{-2} (\lambda_{84} \phi + \mu_{84})] P_{Y_{(2r,2)}} \\ +[(1-(\frac{v}{u})^2)+ \frac{1}{N^2}[r(r-1)(1-(\frac{v}{u})^2) +r u^{-2} (\lambda_{175} \phi + \mu_{175})] P_{Y_{(2r+1,1)}}\} \label{5.49}
\end{multline}
In the limit $x_{12} \rightarrow 0$, $x_{34} \rightarrow 0$, i.e.
\begin{equation} \label{5.50}
u \rightarrow 0, \quad  v \rightarrow 0, \quad Y \rightarrow 0 
\end{equation}
we get
\begin{multline} \label{5.51}
r! \{...\}|_{u=v \rightarrow 0} = (r+1)! \{(1+\frac{r(r+1)}{N^2}) P_{Y_{(2r+2)}} 
+(1+\frac{1}{N^2}\frac{r(5r-11)}{3(r+1)} P_{Y_{(2r,2)}} \\
+\frac{2Y}{r+1}(1+\frac{r(r-1)}{N^2})  P_{Y_{(2r+1,1)}} + \mathcal{O}(Y^2) \}
\end{multline}

From the first term $(P_{Y_{(2r+2)}})$ we see, that Skiba's theorem can be derived by recursion: The two-point function of the normal product
\begin{equation} \label{5.52}
:P_{i_1 i_2...i_{2r}}^{I_1 I_2...I_r} \, \prod_{a=1}^r O^{I_a}(x_a):
\end{equation}
has the free-field term up to order $\mathcal{O}(\frac{1}{N^2})$ included. In all three terms there is no $log\, v$ at dimension $2r+2+l$ and space-time tensor rank $l$. 

\section{Anomalous dimensions}
\setcounter{equation}{0}

We summarize the results of the preceding sections as follows.
\begin{enumerate}
\item
For the $SO(6)$ representations $Y_{(2r+2)}$ and space-time tensors of rank $l$ (all $l \geq 0$), the ''first tower'' with approximate dimensions $2r+2+l$ and ''second tower'' with approximate dimensions $2r+4+l$ obtains no anomalous dimensions due to the factor $v^2$ contained in (\ref{3.37}). But the ''third tower'' with approximate dimensions $2r+6+l$ has non-vanishing  anomalous dimensions. Moreover the scalar $(l = 0)$ field in the first tower has an unrenormalized two-point function (Skiba's theorem).
\item
For the $SO(6)$ representations $Y_{(2r,2)}$ and space-time tensors of rank $l$ (all $l \geq 0$), the ''first tower'' with approximate dimensions $2r+2+l$ does not have anomalous dimensions but the ''second tower'' with approximate dimensions $2r+4+l$ does. This is due to the factor $v$ in (\ref{3.36}).
\item
For the $SO(6)$ representations $Y_{(2r+1,1)}$ and space-time tensors of rank $l$ (all $l \geq 1$), the ''first tower'' with approximate dimensions $2r+2+l$ does not have anomalous dimensions but the ''second tower'' with approximate dimensions $2r+4+l$ does, due to the factor $v$ in (\ref{3.38}).
\end{enumerate}

Whereas the first tower of $Y_{(2r+2)}$ with dimension $\Delta= (2r+2)+2+l$ is known to consist of ($\frac{1}{2}$-BPS) short multiplets, the vanishing of the anomalous dimensions of the second tower, $\Delta=(2r+2)+4+l$ with even $l$, could not be derived yet from any non-renormalization theorem \cite{arut3}. The $(r+1)$-trace fields $Y_{(2r,2)}$ with dimensions $\Delta=(2r+2)+l$ are interpreted as $\frac{1}{4}$-BPS multiplets.
  
We expect that for arbitrary tensor rank $l$, there may exist many different quasi-primary fields with a given $SO(6)$ representation and tensor rank $l$, which differ only by their anomalous dimensions at $\mathcal{O}(\frac{1}{N^2})$ (or higher).
A general analysis shows that the number of such ''almost degenerate'' quasi-primary fields increases typically like $p(l)$ (the number of partitions of $l$). A simple explanation is that the number of fusion channels leading to the same representation increases in this fashion, whereas here we considered only one channel, ending at $Y_{(2r)}$ and $l = 0$ before the last OPE step was performed.

For a resolution of almost degeneracy, one has first to include the full internal symmetry (in this case: $\mathcal{N} = 4$ supersymmetry) and then apply a cohomology argument to discriminate the quasi-primary fields from the derivative fields. This was shown to work in \cite{lang1}, but is out of reach in the present investigation.

We start the harmonic analysis of the leading terms in (\ref{5.51}), namely 
\begin{align}
&P_{Y_{(2r+2)}}+ P_{Y_{(2r,2)}} \,: \, 1+r(\frac{v}{u})^{2} = \sum_{k=0}^\infty A_k (1-V)^k, \label{6.1} \\
&P_{Y_{(2r+1,1)}} \, : \; 1-(\frac{v}{u})^{2} = \sum_{k=0}^\infty B_k (1-V)^k, \qquad (V=\frac{u}{v}) \label{6.2}
\end{align}
which gives
\begin{align}
&A_k = \delta_{k,0}+r(k+1), \label{6.3} \\
&B_k = \delta_{k,0} - (k+1) \qquad (=0 \; \text{for} \; k=0) \label{6.4}
\end{align}
We compare the sums (\ref{6.1}), (\ref{6.2}) with the sum over the ''first tower'' of quasi-primary fields of dimension $\lambda +l$\footnote{$\lambda = 2r+2$ is used as a shorthand in this section.}, obtained from the fusion
\begin{equation*}
(\lambda -2) + (2) \longrightarrow (\lambda +l) 
\end{equation*}
Correspondingly we call the ''second tower'' (''third tower'') of quasi-primary fields those of dimension $\lambda+l+2$ ($\lambda+l+4$), obtained from the fusion
\begin{equation*}
(\lambda -2) + (2) \longrightarrow (\lambda +l+2), \qquad ((\lambda -2) + (2) \longrightarrow (\lambda +l+4)).
\end{equation*}
We apply the formulas of \cite{lang2} for the full exchange amplitude of a quasi-primary field including all derivatives. 

The expression (\ref{6.1})  is expanded correspondingly in contributions of each tower as 
\begin{equation} \label{6.5}
\sum_{l=0}^\infty \gamma^{I}_l \sum_{n=0}^\infty \frac{u^n}{n!} G^I_n(\lambda,l;V) + u \sum_{l=0}^\infty \gamma^{({I \negthickspace I})}_l \sum_{n=0}^\infty \frac{u^n}{n!} G^{I \negthickspace I}_n(\lambda,l;V) + \, \text{contributions of higher towers}
\end{equation}
Setting $u=0$ in (\ref{6.1}) we obtain an equation which allows us to determine the ''coupling constants squared'' $\gamma^I_l$. For this purpose we use
\begin{equation} \label{6.6}
G^I_0(\lambda,l;V) = (V-1)^l \frac{\Gamma(l+2)\Gamma(\lambda+l-2)}{\Gamma(\lambda+2l)} {_2F_1}(l+2, \lambda+l-2; \lambda+2l; 1-V)
\end{equation}
and later on we will also use
\begin{equation} \label{6.7}
G^{I \negthickspace I}_0(\lambda,l;V) = (V-1)^l \frac{\Gamma(l+3)\Gamma(\lambda+l-1)}{\Gamma(\lambda+2l+2)} {_2F_1}(l+3, \lambda+l-1; \lambda+2l+2; 1-V)
\end{equation}
which allows us to guess the function $G_0$ for all towers.

Expanding (\ref{6.6}) in powers of $1-V$ and comparing with (\ref{6.1}) we get the equations
\begin{equation} \label{6.8}
 A_k = \sum_{l=0}^k (-1)^l \frac{\Gamma(k+2)\Gamma(\lambda +k-2)}{(k-l)! \, \Gamma(\lambda +l+k)} \gamma^{I}_l
\end{equation}
It is useful, however, to solve the ''basic'' equation (\ref{6.8}), where $A_k$ is replaced by $\delta_{k,n}$.
The solution of this basic equation is given by 
\begin{equation} \label{6.9}
\gamma^{(n) I}_l = (-1)^n (\lambda +2l-1)\frac{\Gamma(\lambda +l+n-1)}{(l-n)!(n+1)!\,  \Gamma(\lambda +n-2)}
\end{equation}
The basic equation for the second tower is
\begin{equation} \label{6.10} 
\delta_{k,n} = \sum_{l=0}^k (-1)^l \frac{\Gamma(k+3)\Gamma(\lambda +k-1)}{(k-l)!\, \Gamma(\lambda +l+k+2)} \gamma^{(n) I \negthickspace I}_l
\end{equation}
and is solved by 
\begin{equation} \label{6.11}
\gamma^{(n) I \negthickspace I}_l = (-1)^n (\lambda +2l+1)\frac{\Gamma(\lambda +l+n+1)}{(l-n)! (n+2)! \,\Gamma(\lambda +n-1)}
\end{equation}
From the solution (\ref{6.9}) we obtain
\begin{equation} \label{6.12} 
\gamma^{I}_l = \sum_n A_n \gamma^{(n) I}_l=\gamma^{I}_l(A)  = (\lambda +2l-1) \frac{\Gamma(\lambda +l-1)}{l! (n+1)! \, \Gamma(\lambda-2)}[1+\frac{1}{2}(-1)^l \frac{(l+1)!}{(\lambda-1)_{l-1}}] 
\end{equation}
and 
\begin{equation} \label{6.13}
\gamma^{I}_l(B) = \sum_n B_n \gamma^{(n) I}_l = (\lambda +2l-1) \frac{\Gamma(\lambda +l-1)}{l! \, \Gamma(\lambda-2)} [1-(-1)^l \frac{(l+1)^l}{(\lambda-2)_l}]
\end{equation} 
In particular we obtain
\begin{align}
&\gamma^I_0(A) = \frac{1}{2}(\lambda-2)_3, \label{6.14} \\
&\gamma^I_1(A) = 0, \label{6.15} \\
&\gamma^I_2(A) = \frac{1}{2}\lambda(\lambda+3)(\lambda^2-4), \label{6.16}  
\end{align}
and 
\begin{align}
&\gamma^I_0(B) = 0, \label{6.17} \\
&\gamma^I_1(A) = \lambda(\lambda^2-1), \label{6.18} \\
&\gamma^I_2(A) = \frac{1}{2}(\lambda-4)\lambda(\lambda+1)(\lambda+3), \label{6.19}  
\end{align}

In the next step we evaluate the function $G_1^I$. It can be expanded in terms of the coefficients $\tilde{A}_{nm}$, defined in \cite{lang2}, eqs. (2.15), (2.18), (2.23), (2.27). Namely\footnote{This formula is the reason we prefer the variable $1-V$ instead of $Y$.}
\begin{equation} \label{6.20}
G^I_1(\lambda,l;V) = \sum_{m=0}^\infty (\tilde{A}_{1m}^{(l)} + \frac{G^{(l)}_{l-2}}{G^{(l)}_{l}} \tilde{A}_{0m}^{(l-2)}) \frac{(1-V)^m}{m!}
\end{equation}
Here we use Chebyshev coefficients
\begin{equation} \label{6.21}
\frac{G^{(l)}_{l-2}}{G^{(l)}_{l}} = 
\begin{cases}
-\frac{1}{4} (l-1), \quad l \geq 2 \notag \\
0, \quad l < 2
\end{cases}
\end{equation}
A general formula for $G^I_1(\lambda,l;V)$ for arbitray $\lambda$ and $l$ is unknown to us. But if we reformulate (\ref{6.20}) as
\begin{equation} \label{6.22}
G^I_1(\lambda,l;V) = \sum_{n=0}^\infty \frac{E_{lm}}{m!} (1-V)^m
\end{equation}
then we can evaluate some of the $E_{lm}$, e.g.
\begin{align}
&E_{00} = \frac{4(\lambda-2)}{\lambda(\lambda+1)(\lambda-1)^2}, \label{6.23} \\
&E_{10} = \frac{2(\lambda-4)^2}{\lambda(\lambda^2-1)(\lambda^2-4)}, \label{6.24} \\ 
&E_{20} = -(\frac{3}{2}) \frac{3\lambda^3-18\lambda^2+29\lambda+10}{(\lambda-2)_6(\lambda-1)}, \label{6.25} 
\end{align}
The coefficients $E_{lm}$ appear in the determination of $\gamma_l^{I \negthickspace I}$
\begin{equation} \label{6.26}
\gamma_l^{I \negthickspace I} = - \sum_{\substack{0 \leq m \leq l \\ 0 \leq k \leq l+2}} \frac{E_{km}}{m!}\gamma_k^{I} \gamma_l^{(m) I \negthickspace I}. 
\end{equation}
The three expressions (\ref{6.23})-(\ref{6.25}) allow us to determine 
\begin{equation} \label{6.27}
\gamma_0^{I \negthickspace I} = - \sum_{0 \leq k \leq 2} E_{k0} \gamma_k^{I} \gamma_0^{(0) I \negthickspace I}
\end{equation}
with
\begin{equation} \label{6.28}
 \gamma_0^{(0) I \negthickspace I} = \frac{1}{2} \lambda(\lambda^2-1)
\end{equation}
and 
\begin{align}
&\sum_{k=0}^2  E_{k0}  \gamma_k^{I}(A) = -\frac{1}{4(\lambda^2-1)(\lambda-1)} P_3(A), \label{6.29} \\
&\sum_{k=0}^2  E_{k0}  \gamma_k^{I}(B) = -\frac{\lambda-4}{4(\lambda-1)^2(\lambda^2-4)} \tilde{P}_3(B), \label{6.30} 
\end{align}
with
\begin{align}
&P_3(A) = \lambda^3-14\lambda^2+23\lambda+62, \label{6.31} \\
&\tilde{P}_3(B)= \lambda^3-6\lambda^2+15\lambda+62 \label{6.32}
\end{align}

In order to derive the anomalous dimensions for the second tower of $Y_{(2r,2)}$ and $Y_{(2r+1,2)}$ and for the third tower of $Y_{(2r+2)}$ we extract the $log \, u$-terms of $\phi$ from (\ref{3.46}) after applying the Euler identity 
\begin{equation} \label{6.33}
_2F_1(3,4;6;Y) = (1-Y)^{-3} {_2F_1}(2,3;6;1-V) \quad \text{with} \; 1-V = \frac{Y}{Y-1}   
\end{equation}
giving
\begin{equation} \label{6.34}
\frac{1}{u^2} \phi(u,v) = log \,u \, [- \frac{48}{(1-Y)} \sum_{m=0}^\infty \frac{(1-V)^m}{m!} \frac{(m+1)! (m+2)!}{(m+5)!}] 
+ \, \text{terms continuous at} \, u=0, 
\end{equation}
This function is multiplied with (see (\ref{3.36})- (\ref{3.38}))
\begin{align}
&\lambda_{84} = \frac{3}{2} u(1-Y)(2-Y) + \mathcal{O}(u^2), \label{6.35} \\
&\lambda_{105}= u^2 (1-Y)^2, \label{6.36} \\
&\lambda_{175}= u Y (1-Y) \label{6.37}
\end{align}
and it results
\begin{equation} \label{6.38}
u^{-3}(u^{-4}) \lambda_D(u,v) \phi(u,v) = log \,  u \sum_{m=0}^\infty C_m^{(D)} (1-V)^m + \; \text{terms continuous at} \; u=0
\end{equation}
with
\begin{align}
&C_m^{(2r+2)} = -8 \frac{(m+1)_2}{(m+4)_2}, \label{6.39} \\
&C_m^{(2r,2)} = -12 \frac{(m+1) (m^2+5m+12)}{(m+3)_3}, \label{6.40} \\
&C_m^{(2r+1,1)} = 8 \frac{(m)_2}{(m+3)_2} = - C_{m-1}^{(2r+2)}, \label{6.41} 
\end{align}
These coefficients are inserted into the system
\begin{align} 
&\sum_{l=0}^\infty \epsilon_l^{(D)} G_0^{I \negthickspace I}(\lambda,l;V)=\sum_{m=0}^\infty C_m^{(D)} (1-V)^m,\qquad D \in \{(2r,2), (2r+1,1)\}, \label{6.42} \\
&\sum_{l=0}^\infty \epsilon_l^{(2r+2)} G_0^{I\negthickspace I \negthickspace I}(\lambda,l;V)=\sum_{m=0}^\infty C_m^{(2r+2)} (1-V)^m \label{6.43}
\end{align}
which can be solved for $\epsilon_l$ by the method of basic solutions
\begin{align} 
&\epsilon_l^{(D)}= (\lambda +2l+1)\sum_{m=0}^l (-1)^m  \frac{\Gamma(\lambda+l+m +1) C_m^{(D)}}{(l-m)! (m+2)! \Gamma(\lambda +m-1)}, \quad (D \, \text{as in} (\ref{6.42})), \label{6.44} \\
&\epsilon_l^{(2r+2)}= (\lambda +2l+3)\sum_{m=0}^l (-1)^m \frac{\Gamma(\lambda+l+m +3) C_m^{(2r+2)}}{(l-m)! (m+3)! \Gamma(\lambda +m)}  \label{6.45}
\end{align}
Note that $C_0^{(2r+1,1)} = 0$, so that by using (\ref{6.41})
\begin{align}
&\epsilon_{l+1}^{(2r+1,1)} = - \epsilon_l^{(2r+2)}, \label{6.46} \\
&\epsilon_{0}^{(2r+1,1)} = 0 \label{6.47}
\end{align}
The summations (\ref{6.44}), (\ref{6.45}) can be performed and the result be expressed by a finite $l$-independent number of terms analytic in $\lambda$ and $l$.

The anomalous dimensions are then given by
\begin{align}
&\delta_l^{(D)} = \frac{2r}{N^2} \frac{\epsilon_l^{(D)}}{\gamma_l^{I \negthickspace I(D)}}, \quad D \in \{(2r,2), (2r+1,1)\}, \label{6.48} \\
&\delta_l^{(2r+2)} = \frac{2r}{N^2} \frac{\epsilon_l^{(2r+2)}}{\gamma_l^{I \negthickspace I\negthickspace I}(A)} \label{6.49}
\end{align}
where
\begin{align}
&\gamma_l^{I \negthickspace I(2r,2)} =  \gamma_l^{I \negthickspace I}(A), \label{6.50} \\
&\gamma_l^{I \negthickspace I(2r+1,1)} =  \gamma_l^{I \negthickspace I}(B), \label{6.51} 
\end{align}
The factor $2$ on the r.h.s. of (\ref{6.48}), (\ref{6.49}) shows up in the expansion
\begin{equation} \label{6.52}
u^{\frac{1}{2}\delta_l}= \frac{1}{2}\delta_l\,  log \, u + \mathcal{O}(\frac{1}{N^2})
\end{equation}
The only anomalous dimensions that we can give, are
\begin{align} 
&\delta_0^{(2r,2)} = \frac{(\lambda-2)}{N^2} \frac{(-\frac{6}{5})\lambda(\lambda^2-1)}{\gamma_0^{I\negthickspace I}(A)}, \label{6.53} \\
&\delta_0^{(2r+1,1)} =0 \label{6.54}
\end{align}
where $\gamma_0^{I\negthickspace I}(A)$ is obtained from (\ref{6.27}) by insertion of (\ref{6.28}), (\ref{6.29})
\begin{appendix}
\section{Projection operators for $SO(6)$ representations in tensor spa-ces $({\bf{R}}_{6})^n_\otimes$ and their contractions}
\setcounter{equation}{0}
An irreducible representation
\begin{equation} \label{A1}
\sigma \in SO(6) \longrightarrow \Lambda_\sigma \in End(({\bf{R}}_{6})^n_\otimes)
\end{equation}
is characterized by a Young tableau $Y_{l_1, l_2}$, where $l_1 \geq l_2$, $l_1+l_2=n$. We are interested in two-row diagrams only and thus avoid the complications of self-duality. Then the projector on this representation ''$Y$'' satisfies
\begin{align} 
\Lambda_\sigma P_{Y} &=  P_{Y} \Lambda_\sigma: \; \text{invariance}, \label{A2} \\
 P^2_{Y} &=  P_{Y}: \; \text{normalization}, \label{A3} \\
e(Y)^T  P_{Y}&=  P_{Y} e(Y)= n(Y) P_{Y}: \; symmetry \label{A4}
\end{align}
where $e(Y)$ is the symmetrizer of $Y$. These properties characterize the projector. Using a vector basis $\{\epsilon_i\}_{i=1}^6$ for $({\bf{R}}_{6})$ and a corresponding tensor notation for $P_Y$, we get
\begin{equation} \label{A5}
P_Y^{i_1...i_n, j_1...j_n} = \frac{1}{n(Y)} e^T(Y) \{\prod_{k=1}^n \delta^{i_k j_k} - \, \text{trace terms}\}
\end{equation}
with
\begin{equation} \label{A6}
e(Y)^2 = n(Y)e(Y) 
\end{equation}
where the trace terms are such that any contraction vanishes
\begin{align}
\delta_{i_a i_b} P_Y^{i_1...i_n, j_1...j_n} &= 0 \notag \\
&= \delta_{j_c j_d} P_Y^{i_1...i_n, j_1...j_n}. \label{A7}					
\end{align}
We are now interested in contractions ascribed to an index set
\begin{equation} \label{A8}
\omega \subset \{1,2,...,n\}, \; \# \omega = r<n
\end{equation}
and defined by
\begin{equation} \label{A9}
 P_Y^{i_1 i_2...i_n, j_1 j_2...j_n} \prod_{k \in \omega} \delta_{i_k j_k}
\end{equation}
If the residual symmetry is described by the (two-row) Young tableau $Y'$, then the remaining operator must be 
\begin{equation} \label{A10}
\frac{\text{dim} \, Y}{\text{dim} \, Y'}  P_{Y'}^{i_1...i_{n-r}, j_1...j_{n-r}}
\end{equation}
since the total contraction 
\begin{equation} \label{A11}
 P_Y^{i_1 i_2...i_n, j_1 j_2...j_n} \prod_{ \text{all} \, k} \delta_{i_k j_k} = \text{dim} \, Y
\end{equation}
is the dimension of the representation $Y$.
This ''contraction theorem'' is applied in the text several times in the elementary cases where
\begin{equation} \label{A12}
\omega \subset \Delta = \{l_2+1, l_2+2,...,l_1\}
\end{equation}
The proof is based on the fact that this contraction satisfies invariance, $Y'$ symmetry and tracelessness and that only the normalization must be adjusted.
\end{appendix}

\section*{Acknowledgements}
A.M. and L.M. would like to thank the German Exchange Service (DAAD) for their financial support.

\end{document}